\documentclass{article}
\usepackage{graphicx} % Required for inserting images
\usepackage[left=2cm, right=3cm, top=2cm]{geometry}
\usepackage[utf8]{inputenc}

\usepackage{fancyhdr}
\usepackage{mwe} % for the example image
\usepackage[utf8]{inputenc} 
\usepackage{biblatex}
\usepackage{amsmath}
\usepackage{amsthm}
\usepackage{amssymb}
\usepackage{comment}
\usepackage{soul, color}
\usepackage{tasks}
\usepackage{float}
\usepackage{wrapfig}
\usepackage{graphicx}
\usepackage[utf8]{inputenc}
 \usepackage[small]{titlesec} %<size> is big, medium, small, or tiny.

\usepackage{hyperref}
\usepackage{lipsum}
\usepackage{footnote}
\usepackage{tablefootnote} % for table footnotes
\makesavenoteenv{table}

\newbox{\bigpicturebox}

\usepackage{array}
\usepackage{makecell}

\usepackage{amsmath}

\numberwithin{equation}{section}

\usepackage[ruled,vlined]{algorithm2e}
\usepackage{subcaption,graphicx}

\usepackage{dsfont}
\usepackage{bm}
\usepackage{graphicx}
\graphicspath{{pictures/}}
\DeclareGraphicsExtensions{.pdf,.png,.jpg,.gif}

\usepackage{stackengine}

\usepackage{makecell}

\usepackage[column=0]{cellspace}
\usepackage{graphicx}% http://ctan.org/pkg/graphicx

\usepackage[shortlabels]{enumitem}

\usepackage{relsize}

\usepackage{lipsum} % for filler text
\usepackage[OT2,T1]{fontenc}
\DeclareSymbolFont{cyrletters}{OT2}{wncyr}{m}{n}
\DeclareMathSymbol{\Sha}{\mathalpha}{cyrletters}{"58}
 \theoremstyle{plain}
 \newtheorem{thm}{Theorem}[section]
 \newtheorem{lem}{Lemma}[section]

 \theoremstyle{definition}
 \newtheorem{defn}{Definition}[section]

\theoremstyle{remark}
\newcommand{\N}{\ensuremath{\mathbb{N}}}
\renewcommand{\L}{\mathcal{L}}

\newcommand{\Z}{\ensuremath{\mathbb{Z}}}
\newcommand{\Zb}{\ensuremath{\bm Z}}
\newcommand{\R}{\ensuremath{\mathbb{R}}}
\newcommand{\IR}{\mathbb{R}}
\newcommand{\V}{\mathcal{V}}

\newcommand{\G}{\mathcal{G}}

\newcommand{\Xb}{\bm X}

\newcommand{\x}{\bm{x}}
\newcommand{\q}{\bm{q}}

\newcommand{\uu}{\bm{u}}
\newcommand{\vv}{\bm{v}}

\newcommand{\A}{\mathcal{A}}
\newcommand{\E}{\mathbb{E}}
\newcommand{\Prob}{\mathbb{P}}

\newcommand{\lam}{\lambda}

\newcommand{\zb}[1]{\ensuremath{\boldsymbol{#1}}}

\newcommand{\diag}{\mathrm{diag}}

\newcommand{\Lb}{\zb{L}}
\newcommand{\elb}{\zb{\ell}}
\newcommand{\alphabet}{\mathcal{A}_{\delta}}

\newcommand{\Ec}{\mathcal{E}}

\newcommand{\f}{\zb f}
\newcommand{\Sb}{\zb S}

\newcommand{\lra}[1]{\left\langle{#1}\right\rangle}

\newcommand{\notwo}[1]{\|{#1}\|_{2}}

\usepackage[dvipsnames]{xcolor}

\title{Low-Bit Quantization of Bandlimited Graph Signals via Iterative Methods}

\author{ \large Felix Krahmer, He Lyu, Rayan Saab, Jinna Qian,  Anna Veselovska,\\ and Rongrong Wang}
\date{ }

\begin{document}
\sloppy

\maketitle

\begin{abstract}
We study the quantization of real-valued bandlimited signals on graphs, focusing on low-bit representations. 
We propose iterative noise-shaping algorithms for quantization, including sampling approaches with and without vertex replacement. 
The methods leverage the spectral properties of the graph Laplacian and exploit graph incoherence to achieve high-fidelity approximations. 
Theoretical guarantees are provided for the random sampling method, and extensive numerical experiments on synthetic and real-world graphs illustrate the efficiency and robustness of the proposed schemes.
\end{abstract}


\section{State of the art and any preliminary works}\label{sec:intro}

Graph signals provide a natural representation of data in many modern applications, including social networks, web information analysis, sensor networks, biological systems, and machine learning. 
Graph Signal Processing (GSP) has emerged as an active area of research that extends classical signal processing tools to signals supported on irregular graph domains by explicitly accounting for the underlying connectivity structure~\cite{shuman2016vertex,pesenson2009variational}. 
Through the eigenstructure of the graph Laplacian, GSP enables frequency-domain interpretations, filtering, and sampling theory for graph-based data, making it possible to exploit smoothness and low-dimensional structure in many practical signals.

Quantization is a fundamental problem in signal processing concerned with representing continuous-valued data using a finite set of values~\cite{gray1998quantization}. 
While quantization theory is well developed for time-series and image signals, extending these techniques to graph-structured data presents new challenges due to the irregular sampling domain and non-uniform spectral geometry. 
In the graph setting, quantizing a signal $\f$ consists of replacing it with a vector $\q$ whose entries belong to a finite alphabet, such that a faithful approximation of $\f$ can be reconstructed from $\q$. 
A key challenge is to design quantization schemes that exploit graph structure and spectral properties, rather than applying scalar quantization independently at each vertex.

Of particular relevance to this work are noise-shaping quantization schemes, which aim to push the quantization error into spectral components that are suppressed by the reconstruction operator~\cite{chou2015noise}. 
Among them, $\Sigma\Delta$ schemes for bandlimited functions and images exploit dependencies between neighboring samples to shape quantization noise toward high frequencies, where it can be effectively filtered out during reconstruction~\cite{gunturk2003one,daubechies2003approximating,lyu2020sigma,krahmer2022enhanced}. 
Noise-shaping techniques have also been successfully applied to compressive sensing measurements~\cite{gunturk2013sobolev,saab2018quantization,krahmer2012root} and, more recently, to post-training quantization of neural networks~\cite{zhang2022post,gunturk2021approximation,maly2022simple}. 
In all these settings, noise-shaping methods significantly outperform naive memoryless quantization that simply rounds each sample independently.

Despite these advances, noise-shaping quantization methods have not been systematically developed for graph-based signals. 
This is largely due to the challenges associated with modeling dependencies induced by irregular graph connectivity and designing algorithms that effectively exploit graph spectral structure. 
Nevertheless, such methods are of strong practical interest, with potential applications including video halftoning on irregular domains~\cite{rehman2010flicker}, graph embedding and visualization~\cite{cui2020adaptive}, and efficient representations for graph neural networks.

Inspired by the success of noise-shaping schemes in classical and learning-based settings, we consider novel iterative noise-shaping quantization techniques for graph signals with bandlimited graph spectra. 
We introduce two algorithmic frameworks: a permutation-based iterative quantization strategy and a randomized sampling method with replacement. 
Both approaches aim to shape the quantization error toward high graph frequencies, where it can be effectively removed by low-pass reconstruction filters. 
For the randomized sampling method, we provide theoretical error bounds in expectation and with high probability, showing explicit dependence on signal bandwidth, graph incoherence, and the number of samples. 
Extensive numerical experiments on synthetic and real-world graphs demonstrate the effectiveness and robustness of the proposed methods.

\paragraph{Relation to Prior Work.}
This paper is an extended version of our conference publication~\cite{krahmer2023quantization}. 
While the conference paper primarily introduced the algorithmic framework for quantizing bandlimited graph signals, the present work provides a theoretical analysis and extends the empirical one. 
In particular, we establish explicit error bounds for the randomized quantization algorithm, and we provide a significantly larger set of numerical experiments that evaluate performance across different graph structures, bandwidths, and quantization parameters.

Our main contributions are summarized as follows:
\begin{itemize}
    \item We investigate noise-shaping quantization algorithms tailored to bandlimited graph signals, explicitly exploiting graph spectral structure.
    \item We provide theoretical performance guarantees for randomized quantization with replacement, revealing favorable decay of reconstruction error with the number of samples and dependence on graph incoherence.
    \item We demonstrate empirical performance on multiple graph datasets, outperforming baseline memoryless quantization methods.
\end{itemize}

\paragraph{Organization of the paper.}
The remainder of the paper is organized as follows. In Section 2, we introduce the graph signal model, review the notion of bandlimitedness on graphs, and formalize the problem of recovering graph signals from quantized and noisy measurements. Section 3 presents the proposed sensing and reconstruction framework, including the design of the sampling strategy and the iterative recovery algorithm, as well as provides theoretical guarantees for the proposed method. 
Section 4 reports numerical experiments on synthetic and real-world graphs, comparing the performance of the considered approach and illustrating the impact of bandwidth rate and incoherence. Finally, Section 6 concludes the paper and discusses directions for future work.
In Section 5, we provide a detailed theoretical analysis of the SSS-R quantization algorithm, deriving error bounds under bandlimitedness of the graph data and its incoherence.

\section{Graph Signal Model and Preliminaries}

We consider an undirected, connected graph $\G=(\V,\Ec,W)$ with no self-loops, where $\V$ is a set of $N$ vertices, with $N$ assumed to be finite but large, $\Ec$ is a set of edges, and $W$ is a weighted adjacency matrix. Let
\[
d_m := \sum_n W_{mn}
\]
be the degree of the $m$th vertex. The corresponding normalized Laplacian matrix is defined as
\[
\L = D^{-1/2}(D-W)D^{-1/2},
\]
where $D=\diag\{d_1,d_2,\dots,d_N\}$ is the diagonal degree matrix. The matrix $\L$ is symmetric positive semidefinite and admits an orthogonal eigen-decomposition with a matrix of eigenvectors given $\Xb=[\x_1,\x_2,\dots,\x_N]$ and corresponding eigenvalues
\[
\lambda(\G)=\diag\{\lam_1,\lam_2,\dots,\lam_N\},
\]
ordered as
\[
\lam_1 \le \lam_2 \le \cdots \le \lam_N .
\]

A signal or function $f:\V\to\IR$ defined on the vertices of the graph can be represented as a vector $\f\in\IR^N$, where the $n$th component represents the function value at vertex $n$. Generalizing the classical Fourier transform, the eigenvectors and eigenvalues of $\L$ provide a spectral interpretation of the graph. The eigenvalues $\{\lam_1,\lam_2,\dots,\lam_N\}$ are referred to as graph frequencies. The eigenvectors of the Laplacian matrix exhibit increasing oscillatory behavior as the magnitude of the graph frequencies increases \cite{shuman2016vertex}.

The \emph{Graph Fourier Transform (GFT)} of a signal $\f$ is defined as
\[
\widehat{\f}=\Xb^T\f, \qquad \widehat{\f}(\lambda_i)=\lra{\f,\x_i},
\]
where again $\Xb$ is the matrix of eigenvectors of the graph Laplacian ordered as above.

A graph signal $\f$ is called \emph{bandlimited} if there exists $r\in\{1,2,\dots,N\}$ such that its GFT has support only in the frequency interval $[0,\lam_r]$, see \cite{pesenson2009variational,shuman2016vertex}. If we denote by $\Xb_r=[\x_1,\dots,\x_r]$ the matrix formed by the first $r$ eigenvectors, then for each $r$-bandlimited signal $\f$ there exists a vector $\bm\alpha\in\R^r$ such that
\[
\f=\Xb_r \bm\alpha .
\]
Without loss of generality, we assume that bandlimited signals are normalized such that
\[
\|\f\|_\infty \le 1 .
\]

The geometry of the graph and the properties of bandlimited subspaces are commonly characterized using graph incoherence \cite{shuman2016vertex}, defined as follows.

\begin{defn}\cite{shuman2016vertex}
For the $r$-dimensional subspace of graph signals spanned by
$\Xb_r=[\x_1,\dots,\x_r]\in\R^{N\times r}$, let
$P_{\Xb_r}=\Xb_r\Xb_r^T$ denote the orthogonal projector onto this subspace.
The incoherence of the subspace is defined as
\begin{equation}\label{eq:incoh-def}
   \mu_r:= \mu(\Xb_r)=\frac{N}{r}\max_{1\le i\le N} \|P_{\Xb_r}e_i\|_2^2 .
\end{equation}
Similarly, the lower incoherence is defined as
\begin{equation}
    \nu_r:=\nu(\Xb_r)=\frac{N}{r}\min_{1\le i\le N} \|P_{\Xb_r}e_i\|_2^2 .
\end{equation}
\end{defn}

It is easy to verify that for any bandwidth $r$, the incoherence parameter satisfies
$\mu(\Xb_r)\in[1,N/r]$.
For several classes of random graphs, it has been shown that sufficiently large graphs have small incoherence with high probability \cite{dekel2011eigenvectors,brooks2013non}, which indicates that the graph Laplacian eigenvectors are well spread.

In this work, we are interested in the quantization of real-valued bandlimited graph signals, that is, in representing such signals using samples from a finite set $\A\subset\R$, referred to as an alphabet. More precisely, given an $r$-bandlimited signal $\f\in\R^N$, our goal is to find quantized samples $\q\in\A^N$ that provide a good approximation of $\f$ under a suitable quality measure.

Motivated by practical considerations, we seek $\q\in\A^N$ such that $\f$ and $\q$ are close under the action of a low-pass graph filter
$\Lb=[\elb_1,\dots,\elb_N]\in\R^{N\times N}$.
We measure quality using the Euclidean norm, i.e.,
$\|\Lb(\f-\q)\|_2$.

We consider quantization using the finite mid-tread alphabet
\begin{equation}\label{alphabet}
    \alphabet=\A_{\delta, K}:=\{\pm k \delta: 0\le k\le K,\; k\in \Z\},
\end{equation}
where $\delta>0$ is the quantization step size, as well as the infinite mid-tread alphabet
\begin{equation}\label{alphabet-inf}
  \A_{\delta, \infty}:=\{\pm k \delta:\; k\in \Z\}.
\end{equation}
For an alphabet $\alphabet$, we define the associated memoryless scalar quantizer
$Q_\delta:\R\to \alphabet$ by
\[
Q_\delta(z):=\arg\min_{x\in \alphabet} |z-x|.
\]
Since the quantized samples $\q$ are designed to approximate $\f$ under low-pass filtering, we refer to
$\f_q=\Lb\q$ as a \emph{quantized representative} of $\f$ and to
$\Lb(\f-\q)$ as the \emph{quantization error}.

\section{Iterative Noise-Shaping Quantization Methods}

\subsection{Permutation-Based Iterative Quantization}
In this section, we present the first class of proposed noise-shaping quantization methods for bandlimited graph signals. The steps of this graph quantization technique are summarized in Algorithm~\ref{alg:full-with-permuations}.

Assume that $\f\in \R^N$ is an $r$-bandlimited graph signal and $\Lb=[\elb_1,\ldots,\elb_N]\in \R^{N\times N}$ is a low-pass graph filter. Ideally, graph signal quantization could be formulated as finding $\q\in \alphabet^N$ that minimizes the distortion between $\f$ and $\q$ under the action of the low-pass filter, i.e., by solving
\begin{equation}
\q^\sharp=\arg\min_{\widehat{\q}\in \alphabet^{N}} \|\Lb(\f-\widehat \q)\|_2 .
\label{eq:opt:problem}
\end{equation}

Unfortunately, the above problem is an instance of integer least squares and is therefore NP-hard in general \cite{hassibi2002expected}. Instead, inspired by recent methods for neural network quantization \cite{lybrand2021greedy, zhang2022post}, we propose iterative greedy methods for selecting the entries of $\q$.

To this end, we fix a random permutation of the graph vertices,
$\V_{perm}=\{k_1,\ldots,k_N\}$.
For this fixed order, we rewrite the quantization error as a linear combination of the columns of the filter,
\[
\Lb(\f-\q)=\sum_{i=1}^N \elb_{k_i} (\f_{k_i}-\q_{k_i}).
\]
After initializing $\q\in \alphabet^{N}$, for the running index $i$, we update the value of $\q_{k_i}$ by choosing
\begin{equation}\label{eq:2nd-loop-min-prob}
\q_{k_i}
= \arg\min_{\widehat{q}\in \alphabet}
\Big\|\sum_{j\not= k_i}\elb_{j} (\f_{j}-\q_{j})
+ \elb_{k_i} (\f_{k_i}-\widehat q)\Big\|_2 .
\end{equation}

The minimizer in \eqref{eq:2nd-loop-min-prob} admits the closed-form solution
\begin{equation}\label{eq:2nd-loop-min-prob-closed-form}
\q_{k_i}
=Q_{\delta}\Big(\f_{k_i}
+\tfrac{\lra{\elb_{k_i},\uu_i }}{\|\elb_{k_i}\|_2^2}\Big),
\end{equation}
where
\[
\uu_i:=\sum_{j\not= k_i}\elb_{j} (\f_{j}-\q_{j})
\]
is the \emph{state vector} associated with the iteration.
We repeat the above update for all $i=1,\ldots,N$, thereby visiting all vertices in the order specified by $\V_{perm}$.
This procedure corresponds to sampling vertices (or equivalently, columns of the low-pass filter) at random without replacement and quantizing according to the sampled order.

After one full pass over all vertices, it is still possible that some entries of $\q$ can be replaced by other alphabet elements and further reduce the quantization error. This is due to the greedy and local nature of the updates, which in general cannot be expected to find a global optimum. To mitigate this issue, we revisit the vertices following the same order $\V_{perm}$ multiple times until $\q$ no longer changes. This corresponds to performing several passes of the quantization procedure, whose total number we denote by $T\in\N$.

In general, the iterative updates converge to a stationary point, which can be interpreted as a local minimum. A good initialization for Algorithm~\ref{alg:full-with-permuations} can therefore help avoid poor local minima. Here, we propose two initialization strategies, inspired by different quantization techniques (see, e.g., \cite{gunturk2003one, lybrand2021greedy, lyu2020sigma, krahmer2022enhanced}).

In the first initialization strategy, labeled \emph{Step-by-Step-Serving} in Algorithm~\ref{alg:init-1st-order}, and using the same vertex order $\V_{perm}$, we construct an initial candidate $\q$ by sequentially minimizing the growth of the filtered error. Specifically, we first choose $\q_{k_1}$ to minimize
$\|\elb_{k_1}(\f_{k_1}-\q_{k_1})\|_2$.
At step $i$, we collect the accumulated error from the previous $i-1$ steps using the state vector
\[
\uu_{i-1}:=\sum_{j=1}^{i-1}\elb_{k_j} (\f_{k_j}-\q_{k_j}),
\]
and select $\q_{k_i}$ by solving
\[
\q_{k_i}
= \arg\min_{\widehat{q}\in \alphabet}
\Big\|\sum_{j=1}^{i-1}\elb_{k_j} (\f_{k_j}-\q_{k_j})
+ \elb_{k_i} (\f_{k_i}-\widehat q)\Big\|_2 .
\]
The minimizer admits a closed-form expression analogous to \eqref{eq:2nd-loop-min-prob-closed-form}, with the state vector updated accordingly. Visiting all $N$ vertices in this manner provides a reasonable initialization for the iterative refinement in \eqref{eq:2nd-loop-min-prob}.

In the second initialization strategy, labeled \emph{Sigma-Delta-Weights} in Algorithm~\ref{alg:init-1st-order}, we traverse the graph using a breadth-first-search (BFS) procedure. We start from a vertex $i$ with maximum degree. At step $k$, we quantize all $k$-hop neighbors of vertex $i$, where vertices are sorted according to the number of already quantized neighbors. When quantizing vertex $j$, we first add to $\f_j$ a weighted combination of the state variables of its quantized neighbors before applying scalar quantization. This strategy is closely related to the weighted $\Sigma\Delta$ approaches proposed in \cite{krahmer2022enhanced, lyu2020sigma}, which also exploit dependencies between neighboring samples.

We use these two initialization strategies as alternative ways to select starting values of $\q$ in Algorithm~\ref{alg:full-with-permuations}. %Our numerical experiments in Section~\ref{sec:num} demonstrate that the two approaches generally lead to almo quantized vectors $\q\in \alphabet^N$. 
The noise-shaping framework presented in this section exhibits good numerical performance, although no theoretical guarantees are currently available.

\begin{algorithm}[!htbp]
\SetAlgoLined
\KwIn{ Low-pass graph filter $\Lb$, graph signal  $\f$,   vertex order $\{k_1, \ldots, k_N\}$,
number of optimization loops $T\in \N$}
\KwOut{Quantized samples  $\q$, reconstruction $\f_{q}$}

%\If{"weighted"==true}{
 \textit{Assign:} \quad $\q \leftarrow$ Algorithm~\ref{alg:init-1st-order}
%}

\For{epoch = 1:T}{
\For{i = 1:N}{
$\uu_i:=\sum_{j\not=k_i}\elb_{j} (\f_{j}-\q_{j})$ \\
$\q_{k_i}= Q_{\delta}\Big( \f_{k_i} +\tfrac{\lra{\elb_{k_i}, \uu_{i}}}{\notwo{\elb_{k_i}}^2}\Big) $
%{\color{red}Should $\uu$ be updated before $\q$? -- Rayan} {\color{blue}{Thank you, yes, It should be updated before, I also changed $\uu_{i-1}$ in the dot product to $\uu_{i}$ -- Anna}}
}
} 
\textit{Assign}
$\f_q=\Lb \q$
\caption{ Graph Noise Shaping with Permutations}
\label{alg:full-with-permuations}
\end{algorithm}

\begin{algorithm}
\SetAlgoLined
\KwIn{ Low-pass graph filter $\Lb$, graph signal  $\f$,  order of vertices $\{k_1, \ldots, k_N\}$, weight matrix $W$, {%\color{blue} 
maximal hop-distance $s$} }

\KwOut{Quantized samples  $\q$}

\If{Step-by-Step-Serving=true}{
%\textbf{Encoding stage:} \\
Assign $\q \leftarrow 0$

\For{i = 1:N}{
$\q_{k_i}= Q_{\delta}\Big( \f_{k_i} +  \tfrac{\lra{\elb_{k_i}, \uu_{i-1}}}{\notwo{\elb_{k_i}}^2}\Big) $ 

$\uu_i:=\sum_{j=1}^{i-1}\elb_{k_j} (\f_{k_j}-\q_{k_j})+ \elb_{k_i} (\f_{k_i}-\q_{k_i}) $

}
}
\If{Sigma-Delta-Weights=true}{
The set $S$  holds already quantized vertices \\
Initialization: $S \leftarrow \{i\}$, where $i$ is a vertex index satisfying $d_i\geq d_n$, for all $n\in [N]$ \\

Assign $\q_i = Q_{\delta}(\f_i)$ and $\uu_i=\f_i-\q_i$; \\
\While{%\color{blue}
|S| < N}{
\For{k = 1: s}{
$T_k \leftarrow$ index set of $k$-hop neighbours~of~$v_i$\\
$P_k \leftarrow$ sort the indices in $T_k$ according to the number of quantized neighbors, i.e., for $j,l \in P_k$, $j$ will be ahead of $l$ if $|\mathcal{N}_{v_l}\cap S|<|\mathcal{N}_{v_j}\cap S|$.

\For{each $j$ in $P_k$}{
$M_j\leftarrow \mathcal{N}_{v_j} \cap S$ \\
$\q_j = Q_{\delta}\Big(\f_j + \frac{\sum_{k\in M_j} W_{j,k} \uu_k}{\sum_{k\in M_j} W_{j,k}}\Big)$ \\
$\uu_j = \frac{\sum_{k\in M_j} W_{j,k} \uu_k}{\sum_{k\in M_j} W_{j,k}} +\f_j -\q_j$ \\
$S \leftarrow S\cup\{j\}$ \\
}
}
}
}
\caption{Initialization Strategies }
\label{alg:init-1st-order}
\end{algorithm}

\begin{algorithm}[!htbp]
\SetAlgoLined
\KwIn{Low-pass graph filter $\Lb$, graph signal $\f$, number of iterations $M$}
\KwOut{Quantized samples $\q \in \widetilde{\alphabet}^N$, reconstruction $\f_q$}

Assign $\widetilde\q \leftarrow 0 \in \R^M$, \, $\uu_0 \leftarrow 0 \in \R^N$

\For{$i=1:M$}{
sample uniformly an index $1\le k_i\le N$ \\
assign $\vv_i = k_i$ \\
$\widetilde\q_i = Q_{\delta}\!\left( \f_{k_i}+ \dfrac{ \lra{\elb_{k_i}, \uu_{i-1}} }{\|\elb_{k_i}\|_2^2} \right)$ \\
$\uu_i = \uu_{i-1}+ \elb_{k_i}(\f_{k_i}-\widetilde\q_i)$
}
Assign $\q_i = \sum_{j\in \sigma(i)} \widetilde \q_j$ \, and \,
$\f_q= \dfrac{N}{M}\, \Lb\, \q$

\caption{Graph Noise Shaping via Step-by-Step Sampling with Replacement}
\label{alg:NlogN-bits}
\end{algorithm}

\subsection{Random Sampling with Replacement}

In this section, we propose an alternative noise-shaping algorithm for bandlimited graph signals based on random vertex sampling with replacement. In most cases, this algorithm achieves performance comparable to Algorithm~\ref{alg:full-with-permuations}. However, unlike the latter, it allows for a quantitative theoretical analysis of the reconstruction error.

Consider an $r$-bandlimited graph signal $\f\in \R^N$ and a low-pass graph filter
$\Lb=[\elb_1,\ldots,\elb_N]\in \R^{N \times N}$, and let $M \in \N$ denote the total number of iterations. To obtain a quantized vector $\q$, at each iteration $1\le i \le M$, we sample a vertex index $k_i$ uniformly at random from the vertex set $\V=\{1,\ldots,N\}$. As in the previous algorithm, we quantize the value $\f_{k_i}$ by selecting a quantization level that minimizes the accumulated filtered error.

Since sampling is performed with replacement, the same vertex may be selected multiple times. Therefore, we introduce an auxiliary vector of quantized samples $\widetilde{\q} \in \alphabet^M$ and an index vector $\vv \in \N^M$ that records the sampled vertices. Specifically, when vertex $k_i$ is selected at iteration $i$, the quantized value is chosen as
\begin{equation}
    \widetilde{\q}_{i}= \arg\min_{\widehat q\in \alphabet}
    \Big\|
    \sum_{j=1}^{i-1}\elb_{k_j} (\f_{k_j}-\widetilde{\q}_{j})
    + \elb_{k_i} (\f_{k_i}-\widehat q)
    \Big\|,
\end{equation}
and we set $\vv_i = k_i$.

At the end of the iteration process, we obtain the vector $\widetilde\q \in \alphabet^M$ of quantized values and the index vector $\vv\in \N^M$ of sampled vertices. Since $M$ may differ from the ambient signal dimension $N$, we aggregate the entries of $\widetilde \q$ corresponding to the same vertex in order to form the final quantized signal $\q \in \R^N$. Specifically, we define
\[
\q_i = \sum_{j\in \sigma(i)} \widetilde \q_j,
\qquad
\sigma(i):=\{k:\, \vv_k=i\}.
\]
This procedure is summarized in Algorithm~\ref{alg:NlogN-bits}. Note that, in general, the resulting vector $\q$ does not belong to $\alphabet^N$, but instead lies in a slightly larger alphabet set denoted by $\widetilde{\alphabet}$. Nevertheless, it can be shown that when $M=O(N\log N)$, the effective alphabet size satisfies
$|\widetilde{\alphabet}|\lesssim |\alphabet|\log N$.

The following result shows that $\f$ can be accurately approximated from the quantized vector $\q \in \widetilde{\alphabet}^N$.

\begin{thm}\label{Thm1}
Consider a bandlimited graph signal $\f$, where $\f=\Xb_r \bm \alpha$ for some
$\bm \alpha\in \R^r$, with $c \le \|\f\|_{\infty}\le 1$. Assume that the parameter $K$ in the definition~\eqref{alphabet} of the alphabet $\mathcal{A}_{\delta, K}$ satisfies $K\delta>1$, and that $\Xb_r$ satisfies the incoherence property~\eqref{eq:incoh-def} with constant $\mu>0$. Let $M>0$ be the number of iterations in Algorithm~\ref{alg:NlogN-bits}, and let $\q \in \widetilde{\alphabet}^N$ be the resulting quantized vector. Defining $\f_q:=\frac{N}{M}\Lb\,\q$, then with probability at least $1-\delta$, the relative quantization error satisfies
\[
    \frac{\|\f_q-\f\|_2^2}{\|\f\|_2^2}
    \le C\,\mu^2\,\frac{r^2\log^2\!\left(\frac{r}{\delta}\right)}{M},
\]
where $C$ is an absolute constant. In addition, $\widetilde{\q}$ can be represented using
$O\!\left(N \log\log \frac{N}{\delta}\right)$ bits.
\end{thm}

\section{Simulation Results}\label{sec:num}

\begin{figure}[!htbp]

\subcaptionbox{ Graph function $\f$ \label{fig:num:1:a}}{\includegraphics[width=2.2in]{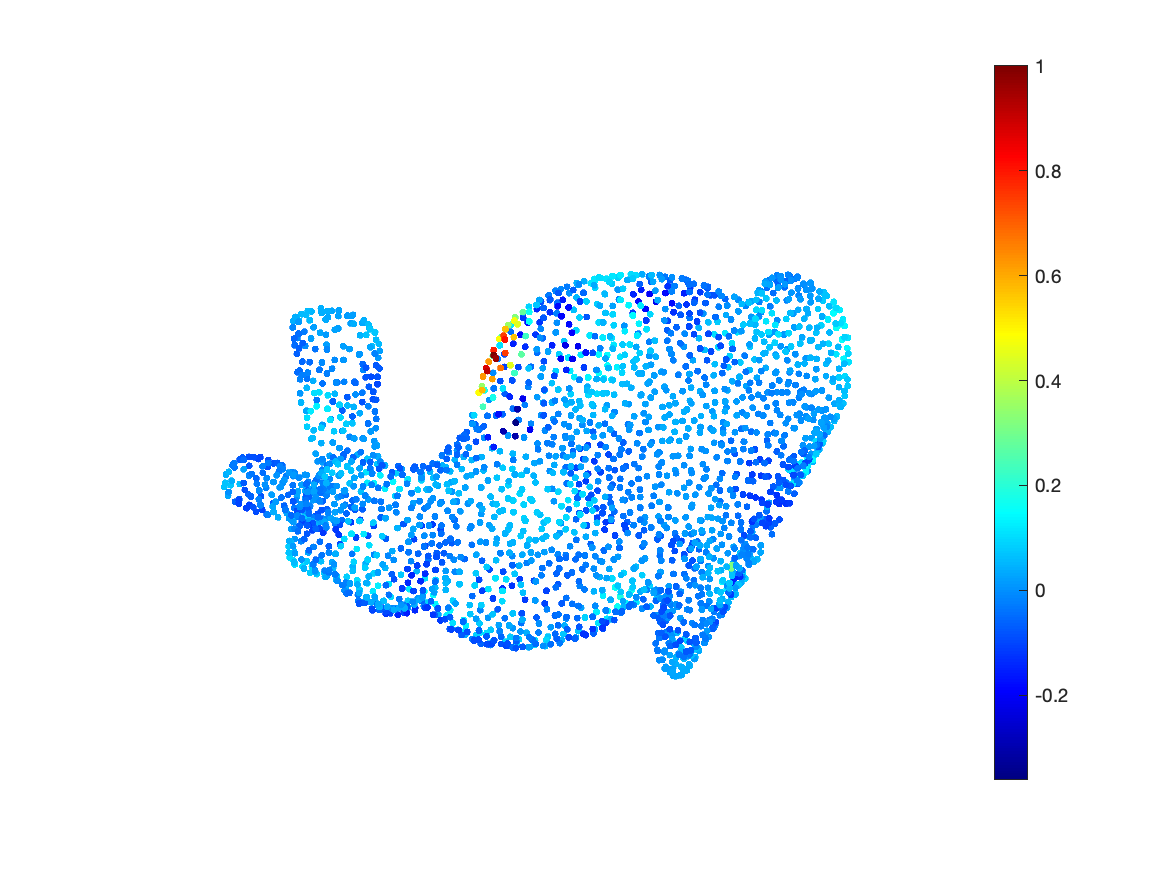}}%{Quant_error_all_1st_schemes_log_1.png}}
\subcaptionbox{ Spectrum of $\f$ \label{fig:num:1:b}}{\includegraphics[width=2.2in]{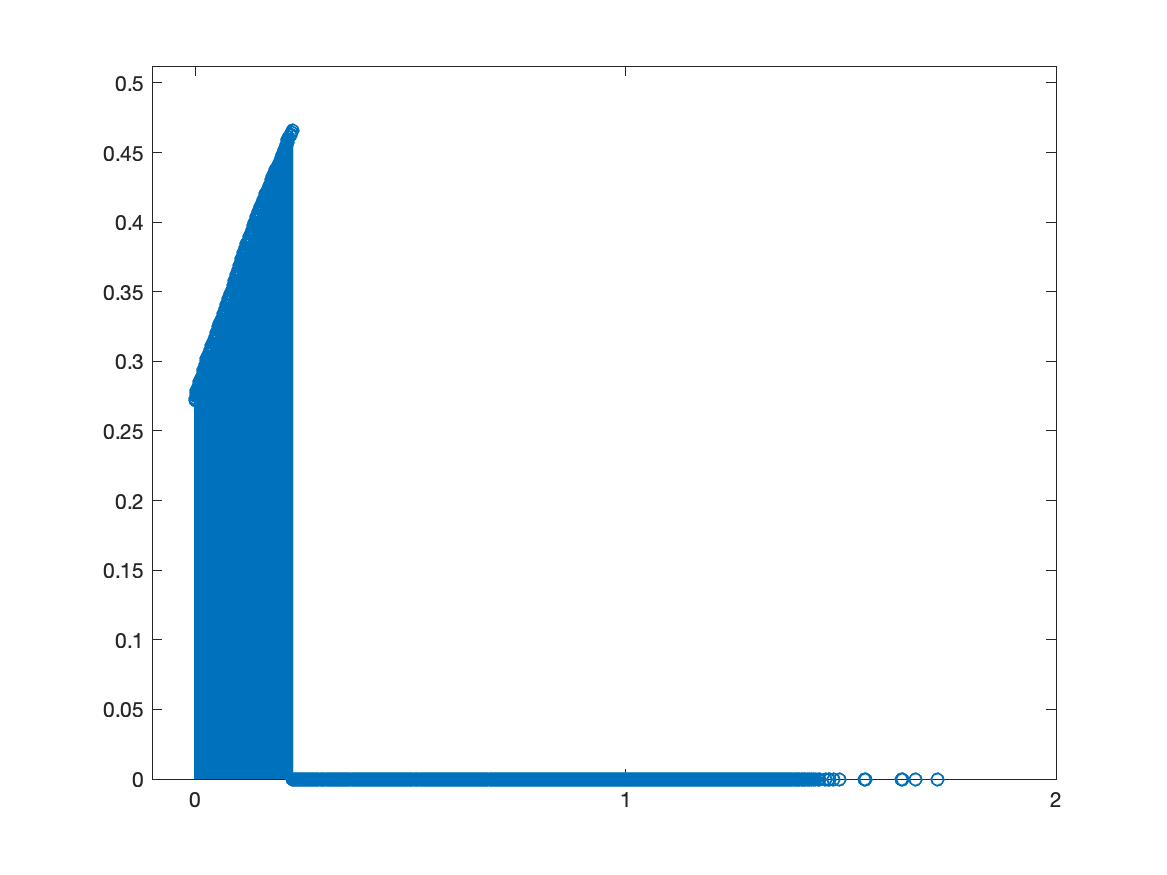}}%{Quant_error_all_1st_schemes_log_1.png}}
\subcaptionbox{  Quant. noise $\f-\f_q$  \label{fig:num:1:c}}{\includegraphics[width=2.2in]{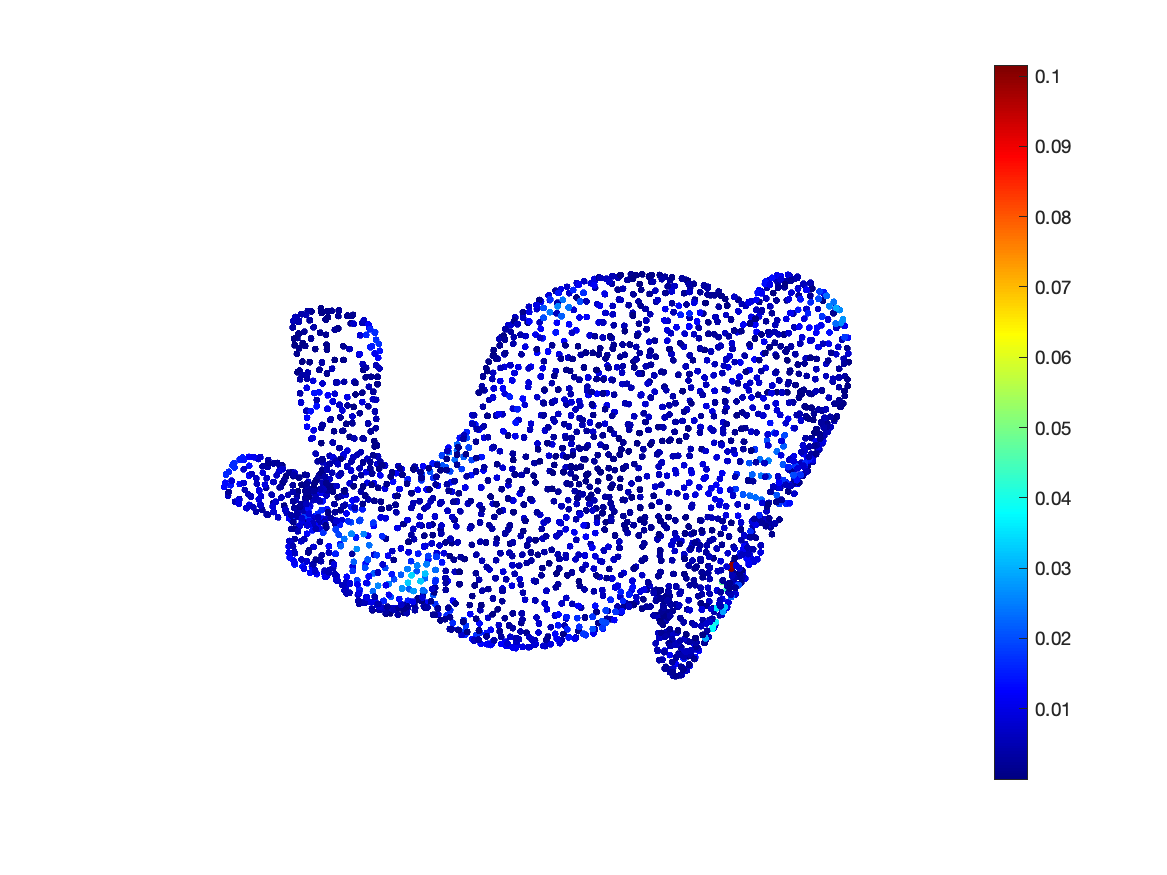}}%

\hskip 5pt

  \subcaptionbox{  Quantized repr. $\f_q$  \label{fig:num:1:d}}{\includegraphics[width=2.2in]{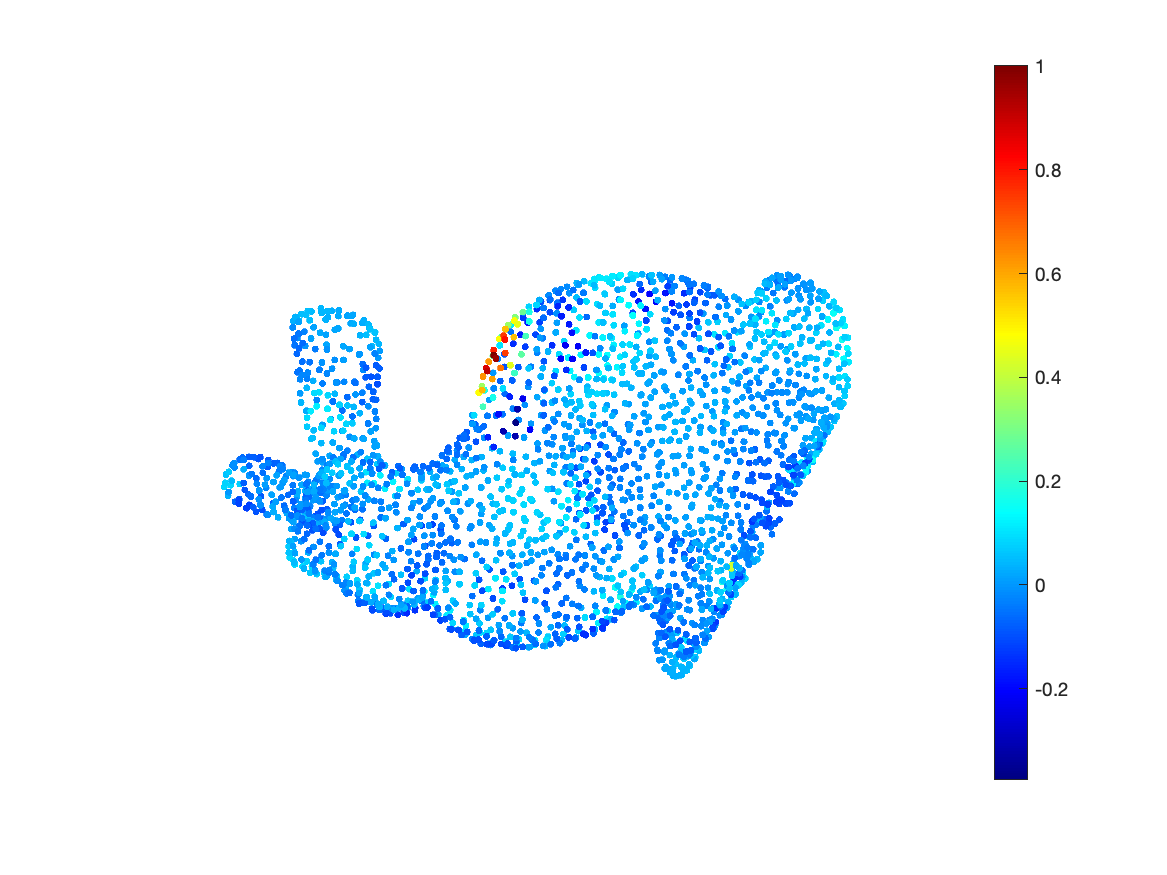}}
  \subcaptionbox{   Spectrum of $\f_q$ \label{fig:num:1:e}}{\includegraphics[width=2.2in]{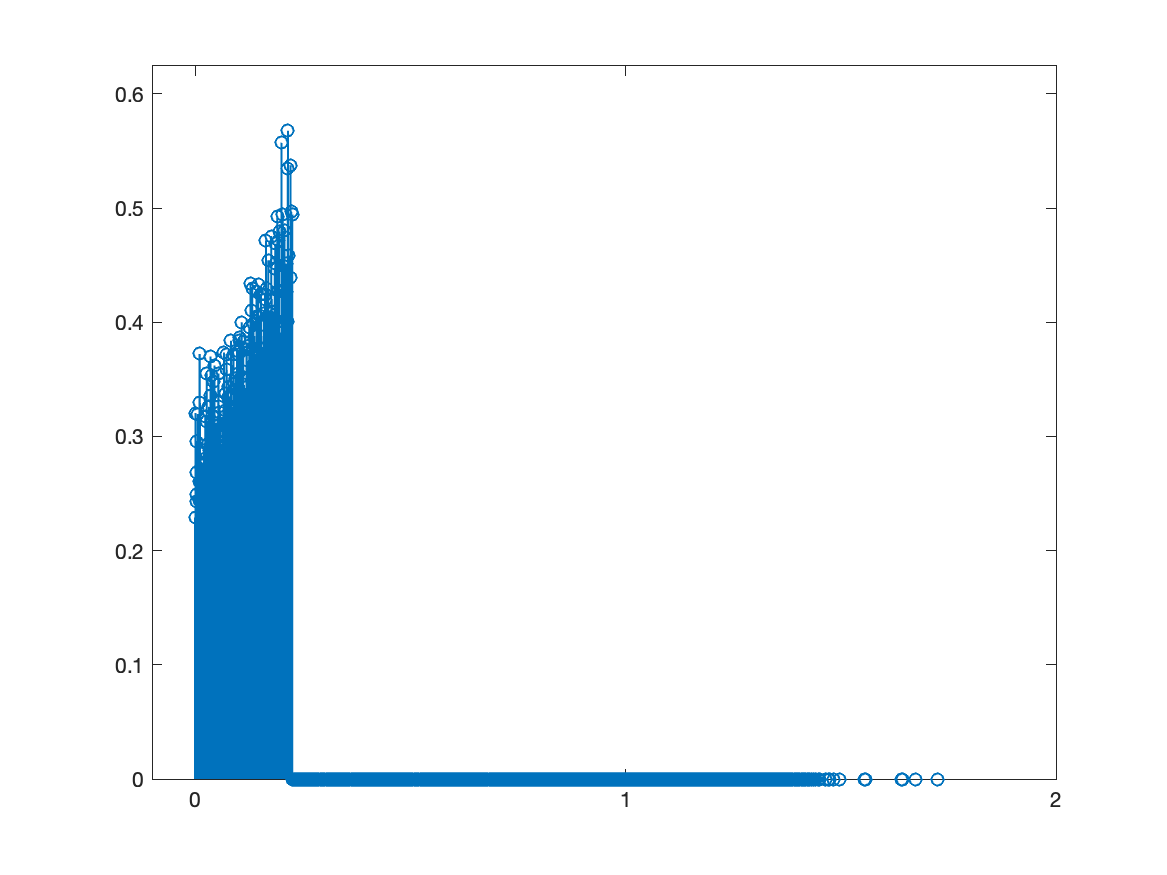}}
  \subcaptionbox{ Spectrum of  $\f-\q$  \label{fig:num:1:f}}{\includegraphics[width=2.2in]{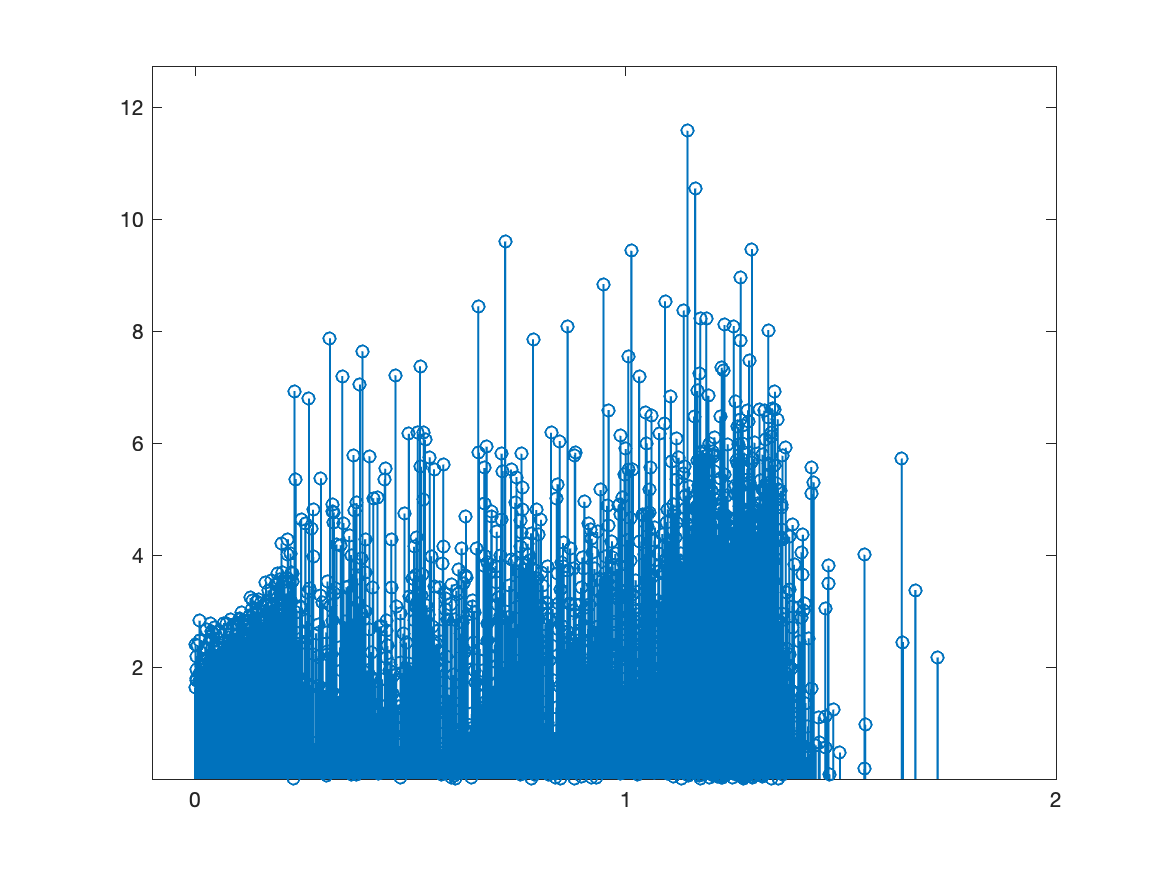}}
  \caption{Performance of the Algorithm~\ref{alg:NlogN-bits} on the bunny graphs of size ${N=2503}$ for a  graph signal $\f$ with bandwidth ${r=100}$. }
  \label{fig:num:1}
\end{figure}

We illustrate the performance of the proposed noise-shaping quantization algorithms on  different graphs and compare them in terms of relative signal reconstruction error.

\paragraph{Experiment 1: Effect of quantization on bandlimited graph signals.}
Figure~\ref{fig:num:1} illustrates the effect of quantization on an $r$-bandlimited graph signal defined on the bunny graph with $N=2503$ nodes and bandwidth $r=100$. Figure~\ref{fig:num:1:a} shows the original graph signal $\f$, which is smooth over the graph and spatially localized. Its graph Fourier spectrum, shown in Figure~\ref{fig:num:1:b}, confirms that most of the signal energy is concentrated in the first $r$ graph frequencies, validating the bandlimited assumption.

Figure~\ref{fig:num:1:c} displays the quantization error $\f - \f_q$, which is spatially dispersed and of relatively small magnitude compared to the signal amplitude. The quantized signal $\f_q$, shown in Figure~\ref{fig:num:1:d}, preserves the main spatial structures of the original signal and  its Fourier. The noise shaping effect is highlighted in Figure~\ref{fig:num:1:f}, which shows that the spectrum of the error $\f - \f_q$ is dominated by high-frequency components, while the low-frequency components remain relatively small.

Overall, the results indicate that after quantization, distortion is mainly confined to higher graph frequencies, while the low-frequency content that captures the smooth structure of the signal is largely preserved. This observation supports the use of low-pass or bandlimited recovery models in the presence of quantization.

\begin{figure}[!htbp]

 \subcaptionbox{ % 
Minnesota graph \label{fig2m:a}}{\includegraphics[width=2.5in]{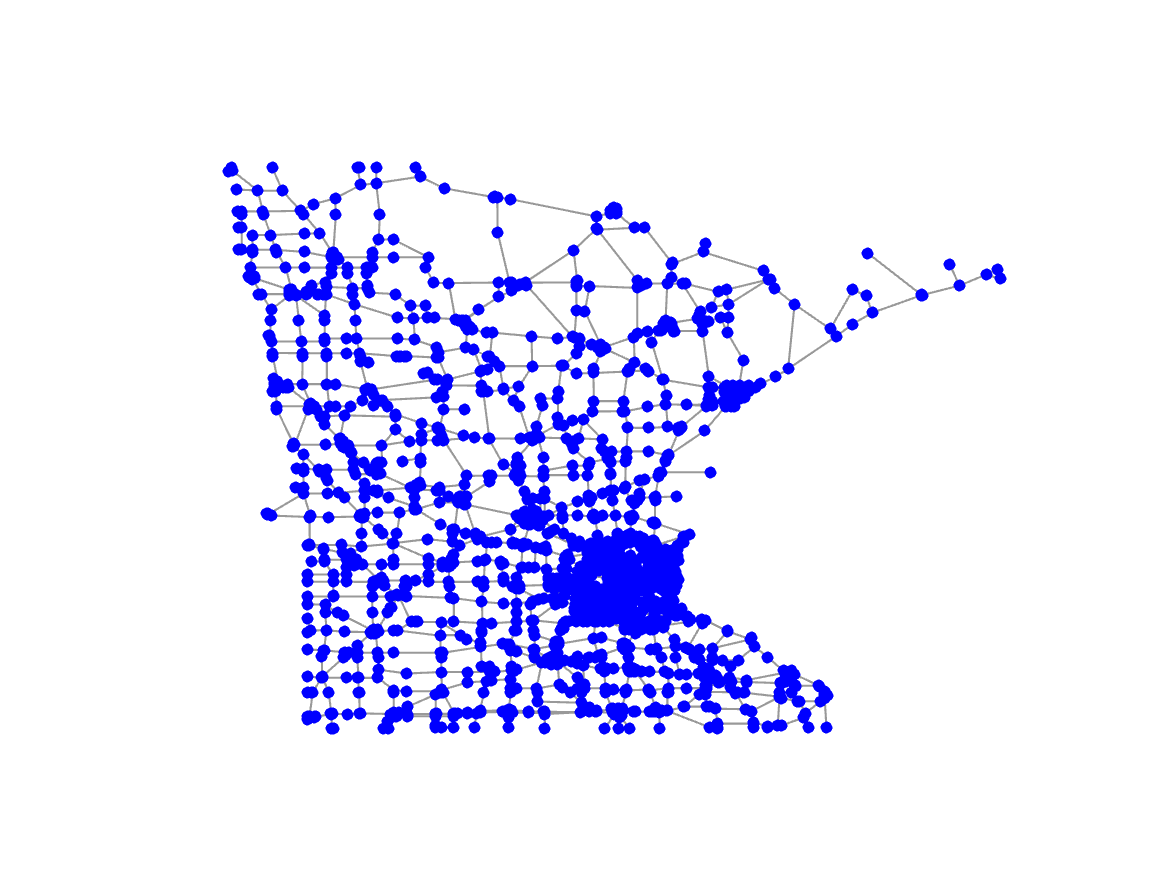}}%
 \hfill%
  \subcaptionbox{%\color{blue}  
Relative error  \label{fig2m:b}}{\includegraphics[width=3in]{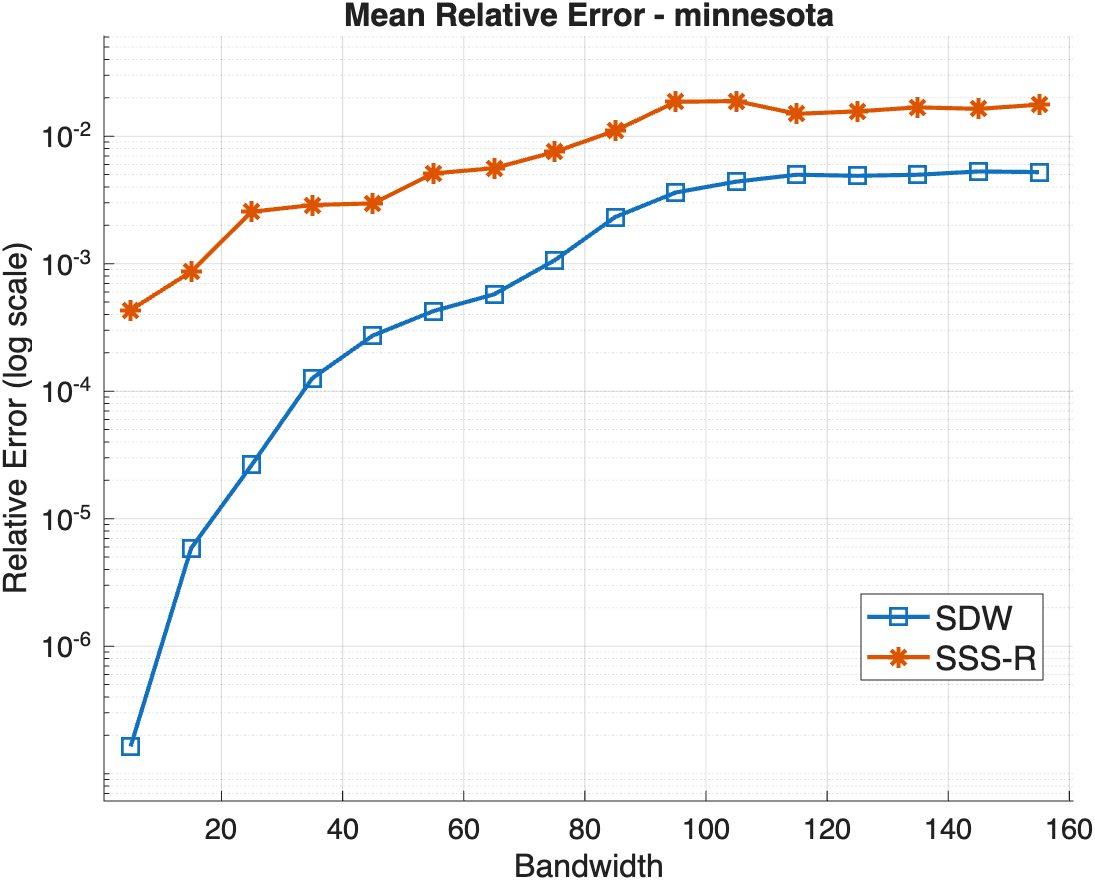}}% 
  \hfill%
%\vspace{2mm}

  \subcaptionbox{ Swiss roll graph \label{fig2m:c}}{\includegraphics[width=2.5in]{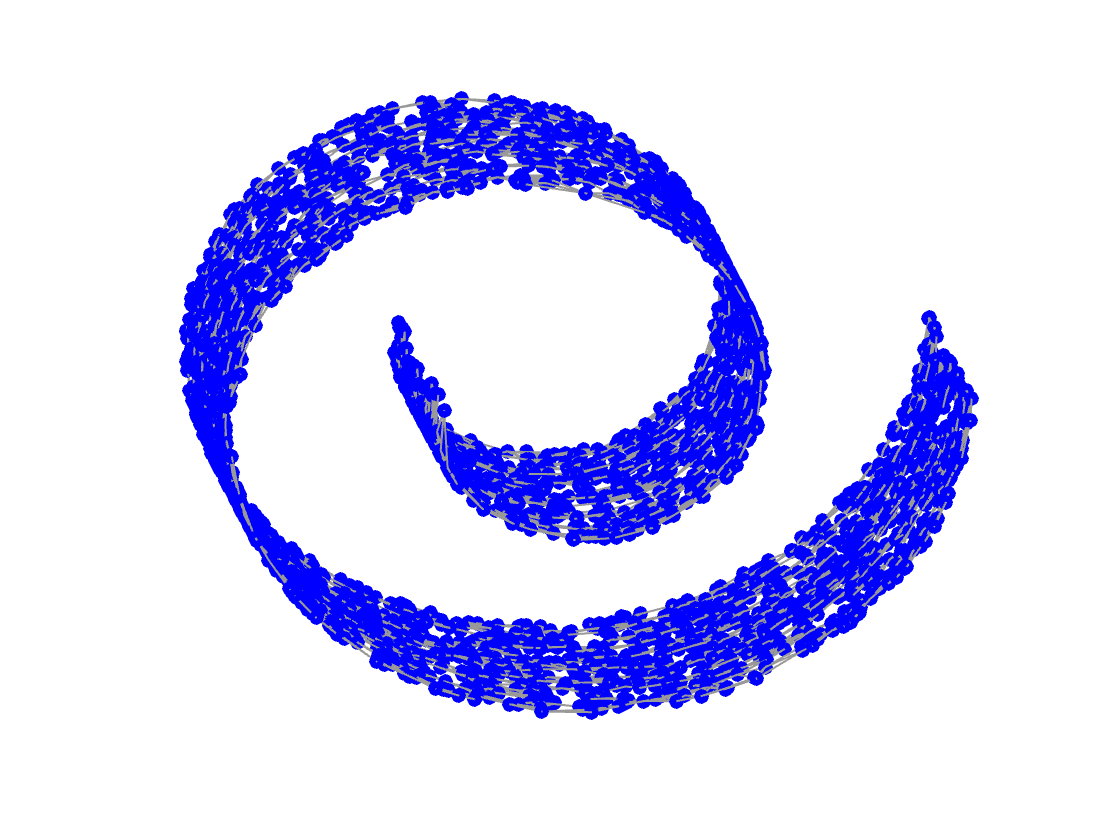}}%
%{Quant_error_all_1st_schemes_log_1.png}}
  \hfill%
  \subcaptionbox{   Relative error \label{fig2:d}}{\includegraphics[width=3in]{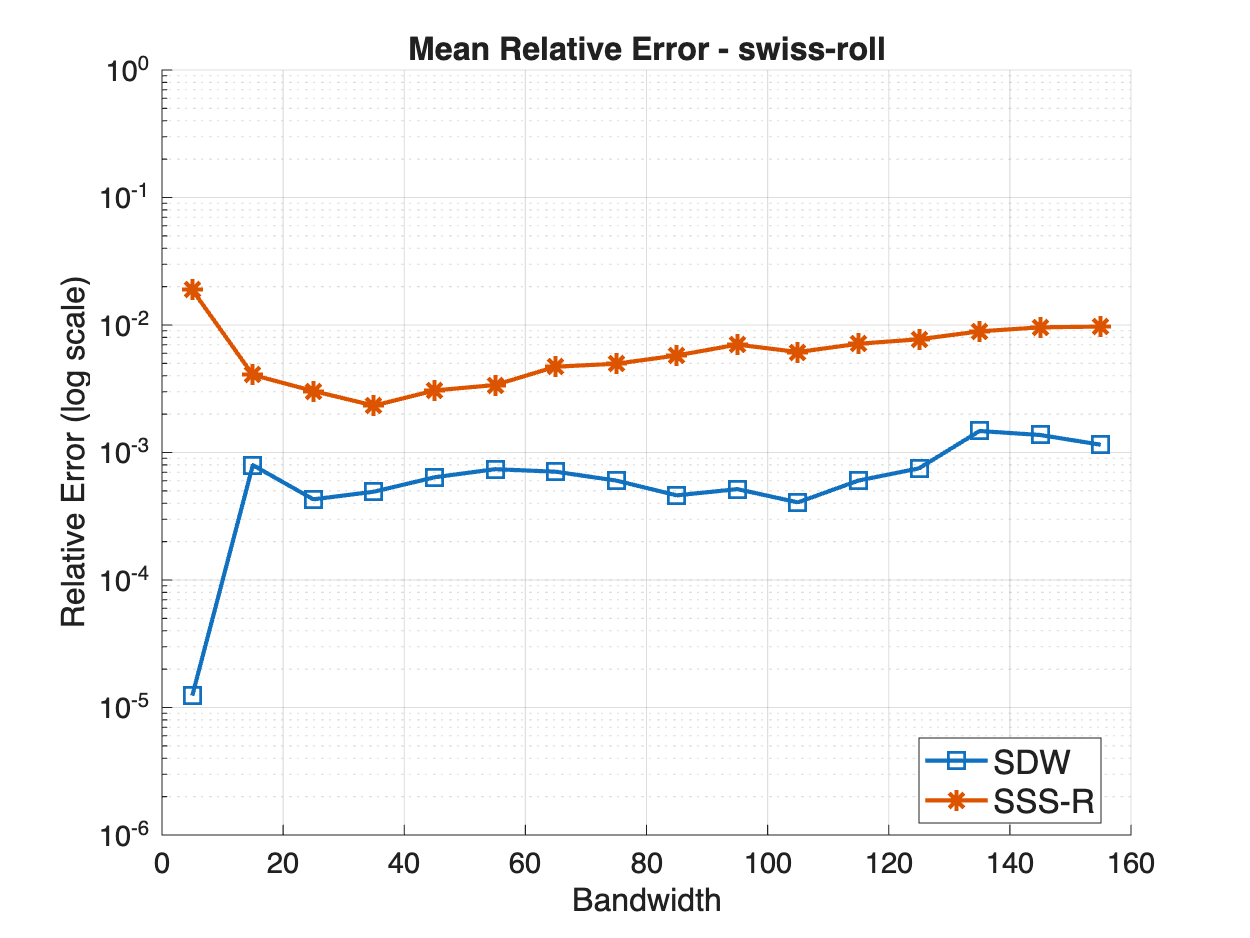}}%

   \subcaptionbox{ % \color{blue}
Bunny graph \label{fig2m:e}}{\includegraphics[width=2.5in]{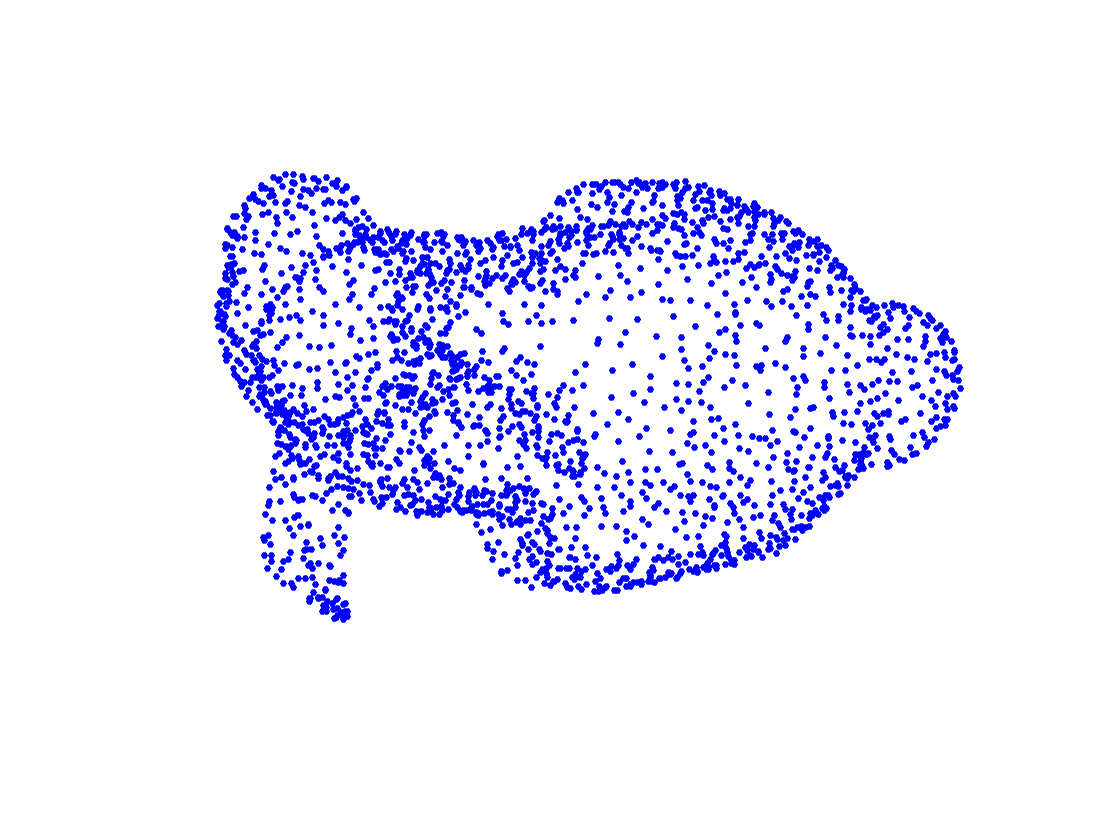}}%
 \hfill%
  \subcaptionbox{%\color{blue}  
Relative error  \label{fig2m:f}}{\includegraphics[width=3in]{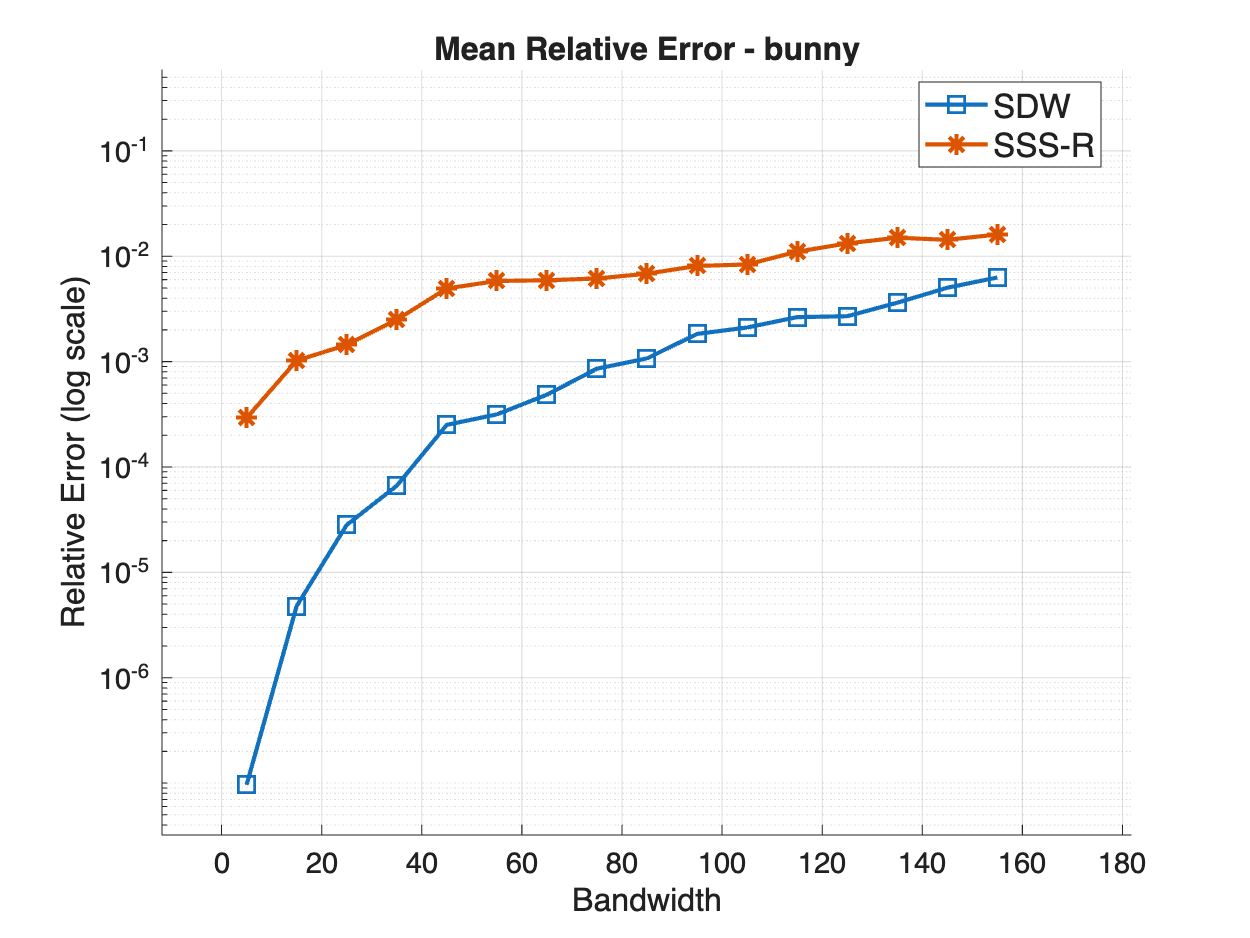}}%

  \caption{Illustration of performance of the proposed  quantization algorithms for  bandlimited graph signals on different graphs:   Algorithm~\ref{alg:full-with-permuations} with the initialization \textit{Sigma-Delta-Weight} (SDW)  and \textit{Step-by-Step-Serving with Replacement} (SSS-R) presented in ~Algorithm~\ref{alg:NlogN-bits}.   }
  \label{fig:num:graphs-and-performance}
\end{figure}

\begin{figure}[!htbp]

  \subcaptionbox{ Sphere graph \label{fig3m:a}}{\includegraphics[width=2.5in]{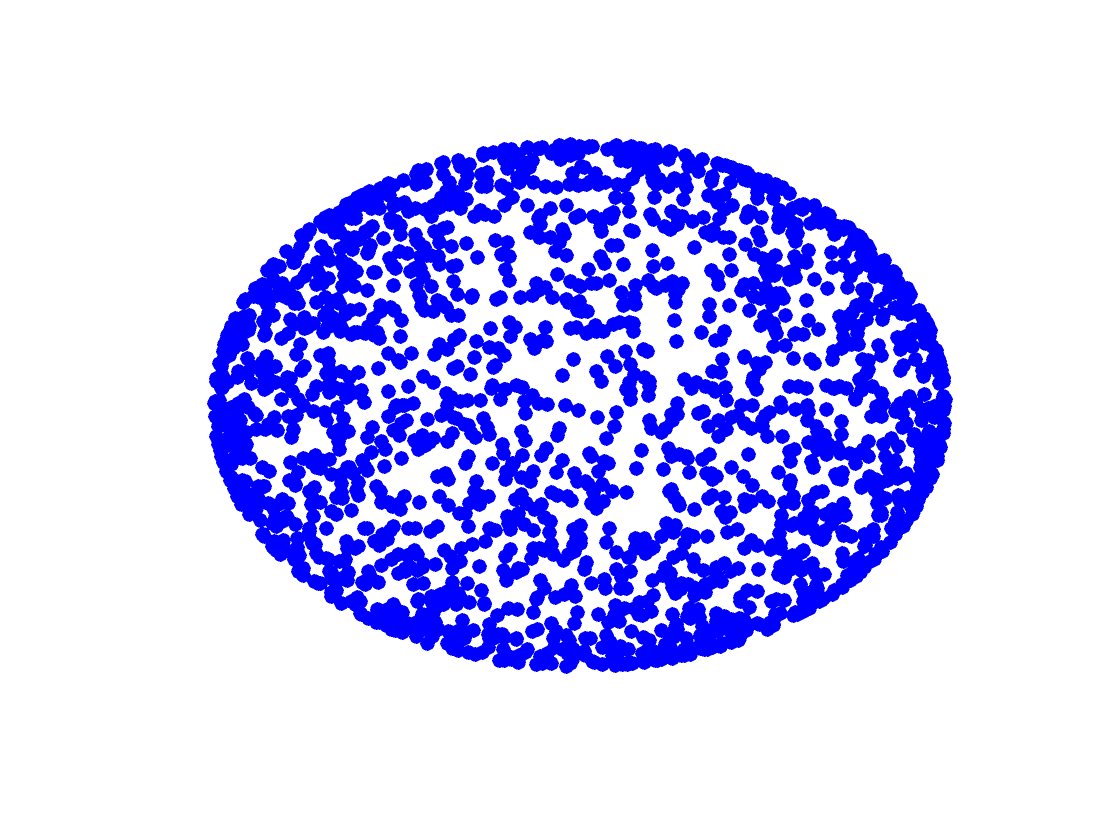}}%
%{Quant_error_all_1st_schemes_log_1.png}}
  \hfill%
  \subcaptionbox{   Relative error \label{fig3m:b}}{\includegraphics[width=3in]{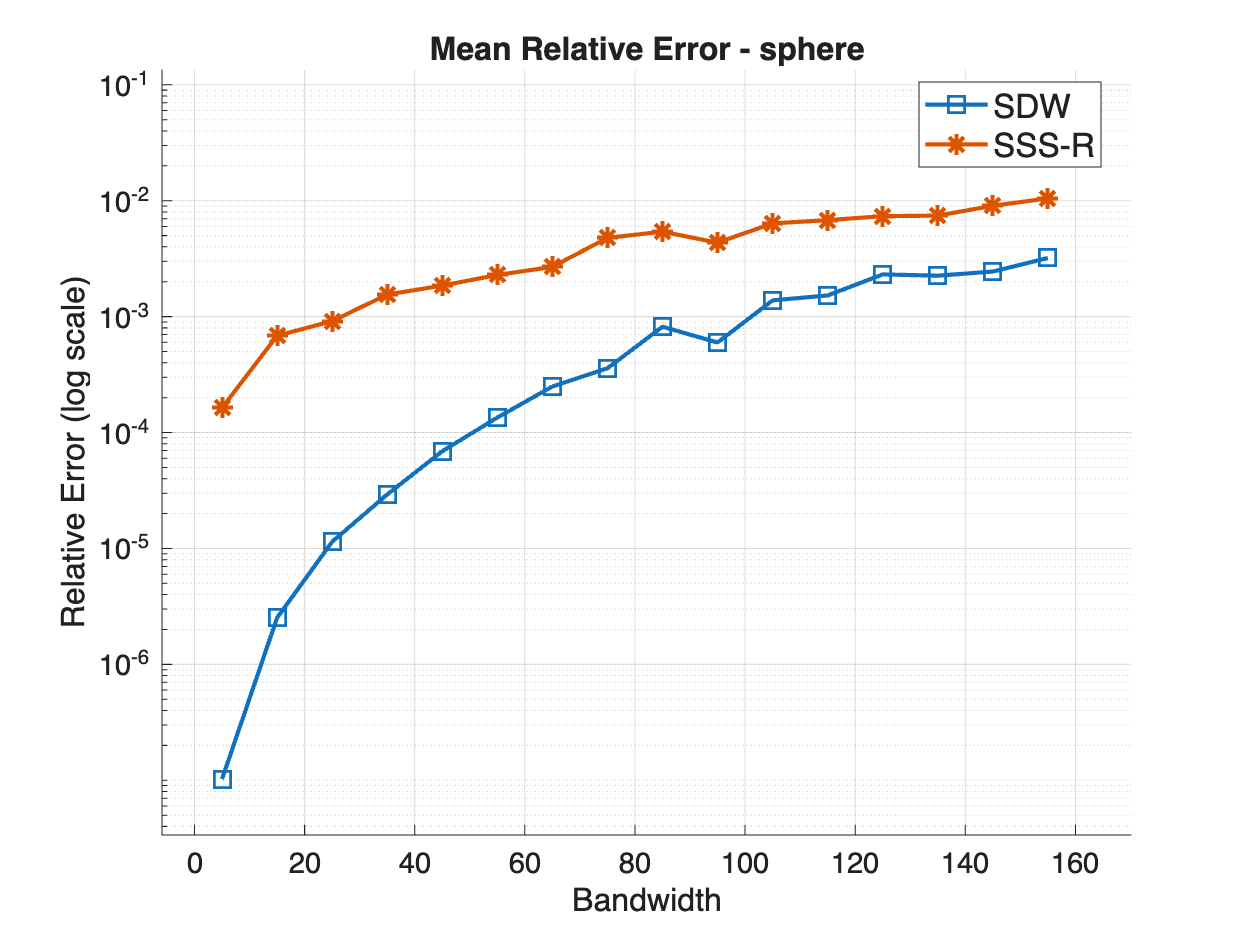}}%

  \subcaptionbox{  2D grid graph \label{fig3m:c}}{\includegraphics[width=2.5in]{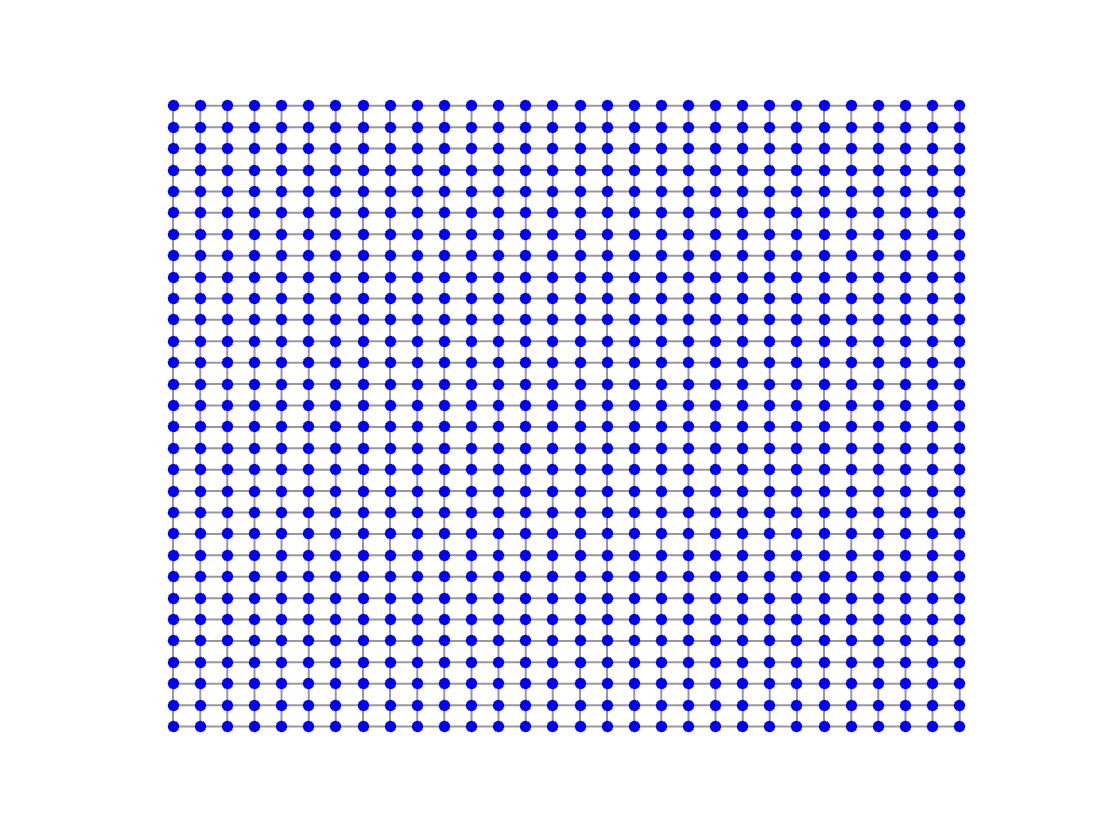}}%{Quant_error_all_1st_schemes_log_1.png}}
 \hfill
  \subcaptionbox{ Relative error \label{fig3m:d}}{\includegraphics[width=3in]{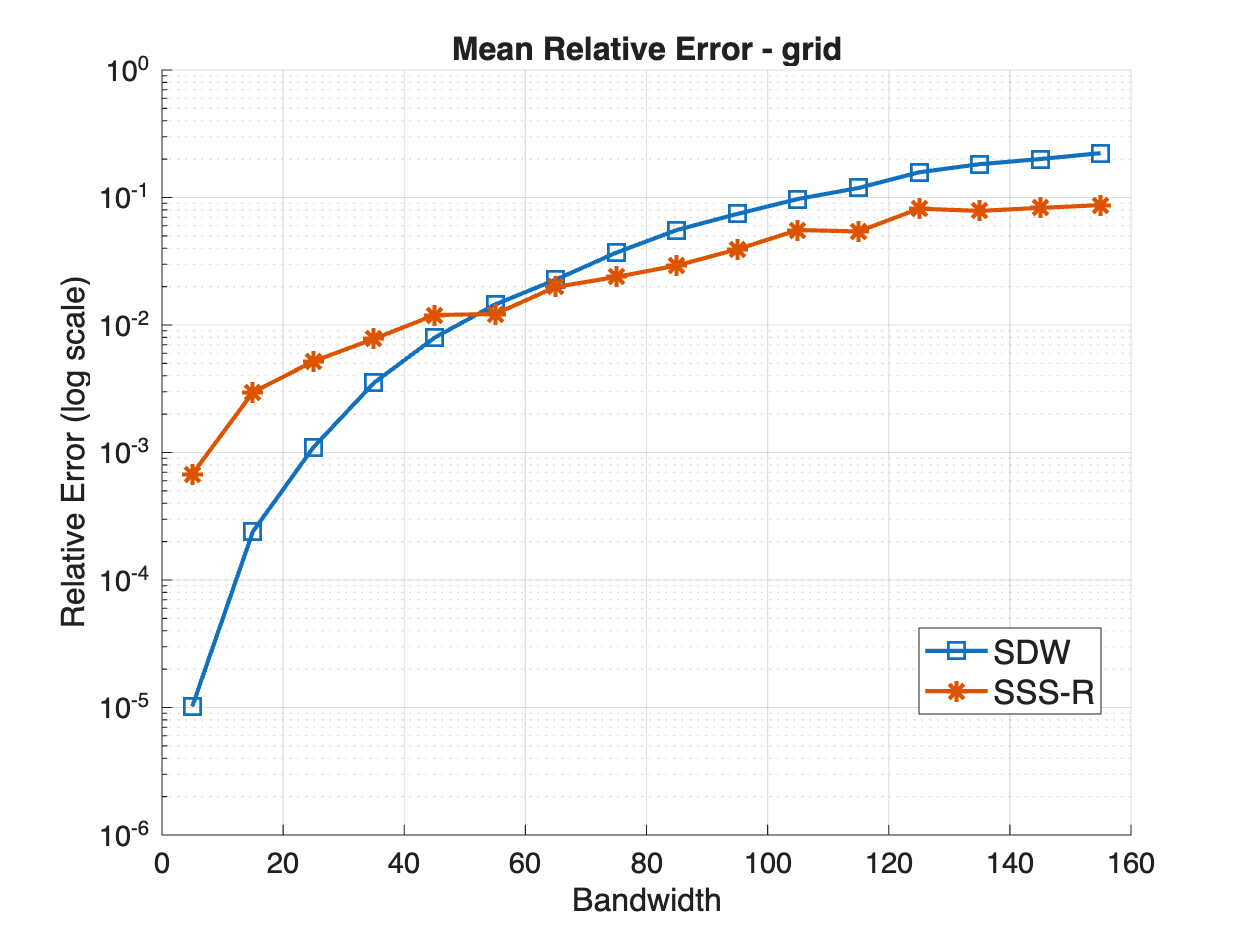}}%

  \caption{Illustration of performance of the proposed  quantization algorithms for  bandlimited graph signals on different graphs:   Algorithm~\ref{alg:full-with-permuations} with the initialization \textit{Sigma-Delta-Weight} (SDW) and \textit{Step-by-Step-Serving with Replacement} (SSS-R) presented in ~Algorithm~\ref{alg:NlogN-bits}. }
  \label{fig:num:graphs-and-performance-2}
\end{figure}

\paragraph{Experiment 2: Performance of the Proposed Quantization Methods Across Diverse Graphs.} We consider a grid graph, %of size %$N=30\times 30$, 
a Swiss roll graph, a bunny graph, a sphere graph, and the Minnesota road map graph shown in Fig.~\ref{fig2m:a}--\ref{fig2m:f} and in Fig.~\ref{fig3m:a}-\ref{fig3m:d}. These graphs exhibit different incoherence properties: the grid graph has relatively low incoherence compared to the bunny graph, while the Swiss roll graph is known to be highly coherent \cite{shuman2016vertex}. For visualization and graph construction, we use the GSPBox toolbox for graph signal processing \cite{perraudin2014gspbox}.

For each graph, we construct $r$-bandlimited signals
\[
\f_r = \Xb_r \bm{\alpha},
\]
with bandwidth $r \in [5,155]$, and normalize each signal such that $\|\f_r\|_\infty = 1$. Each signal is then quantized using two approaches. First, we apply Algorithm~\ref{alg:full-with-permuations} with  initialization Sigma-Delta-Weight (SDW), using an alphabet $\alphabet$ of size approximately $\log\log(N)$ bits and setting $T=10$. Second, we apply Algorithm~\ref{alg:NlogN-bits} with the binary alphabet $\mathcal{A}=\{-1,1\}$ and $M = N\log(N)$ iterations, so that, on average, all vertices are visited, as suggested by the coupon collector problem. We did not observe a substantial difference in performance between the Step-by-Step-Serving (SSS) and SDW initialization strategies; therefore, we use only SDW.
Since in Algorithm~\ref{alg:NlogN-bits} storing each element of $\q$ requires approximately $\log\log(N)$ bits, this places the two algorithmic settings on roughly equal footing.

To compare performance, we measure the relative reconstruction error
\begin{equation}\label{eq:relat-im-error}
    \frac{\|\f_q - \f\|_2^2}{\|\f\|_2^2}.
\end{equation}
The experimental results are shown in Fig.~\ref{fig2m:a}--\ref{fig2m:f} and in Fig.~\ref{fig3m:a}-\ref{fig3m:d}.
As observed, incorporating more graph structure in the initial estimate, as in the SDW initialization, leads to improved overall performance. Moreover, Algorithm~\ref{alg:NlogN-bits} performs better on graphs with lower incoherence, which is consistent with our theoretical results presented in Theorem~\ref{Thm1}.

\begin{figure}[!htbp]

  \subcaptionbox{ 2D grid graph  \label{fig4:a}}{\includegraphics[width=3in]{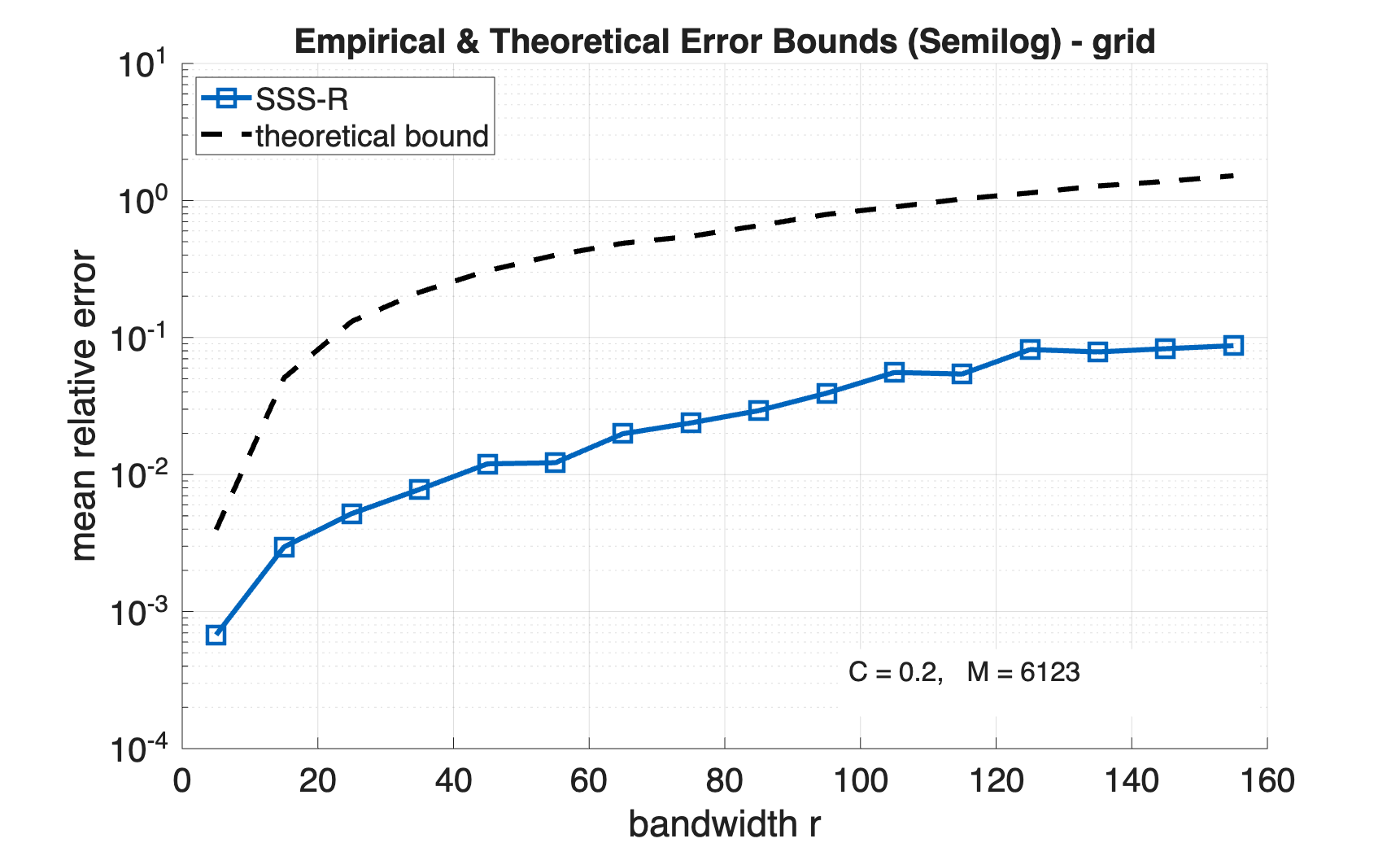}}%
%{Quant_error_all_1st_schemes_log_1.png}}
  \hfill%
  \subcaptionbox{   Bunny graph \label{fig4:b}}{\includegraphics[width=3.2in]{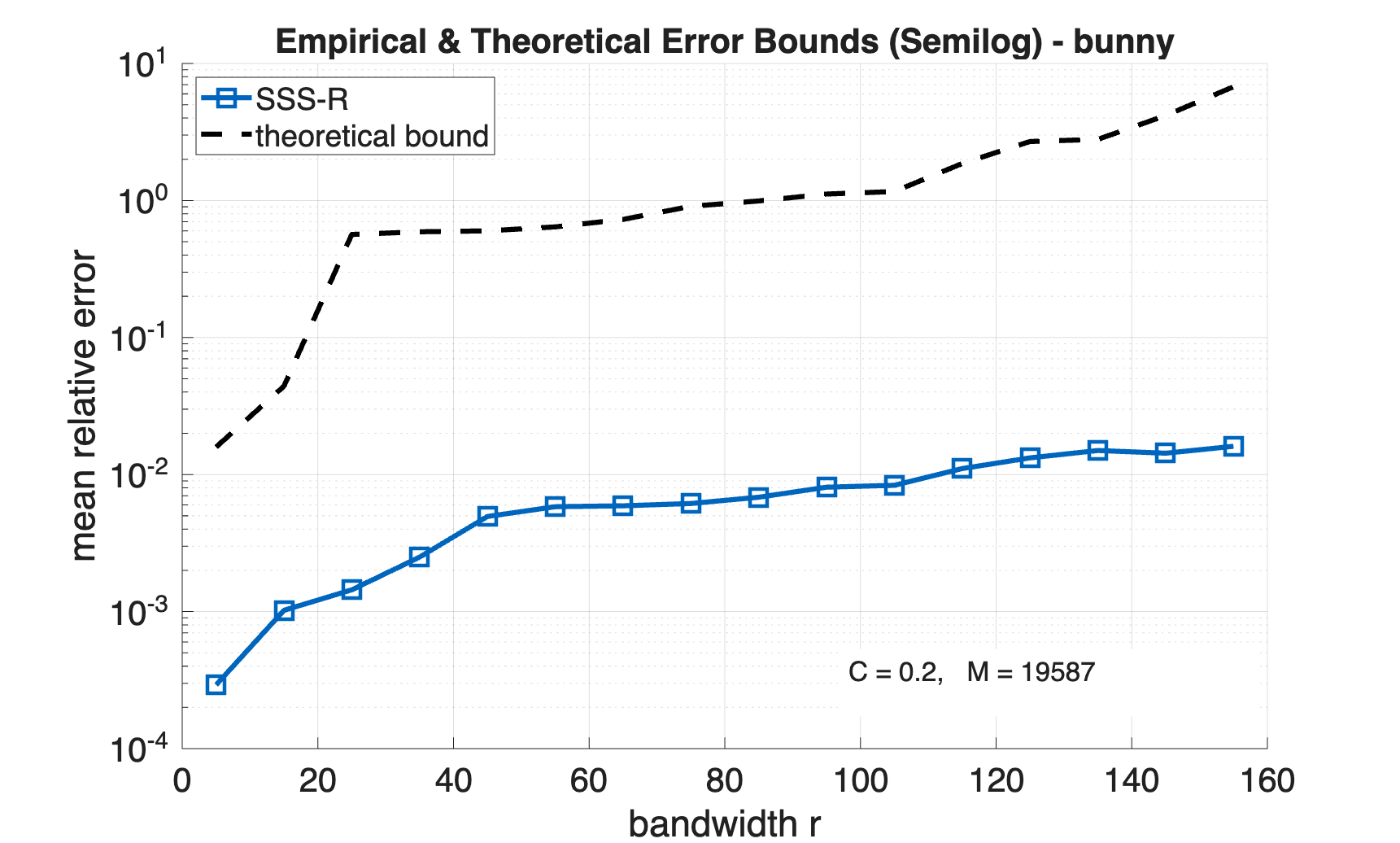}}%

  \subcaptionbox{  Sphere graph \label{fig4:c}}{\includegraphics[width=3.2in]{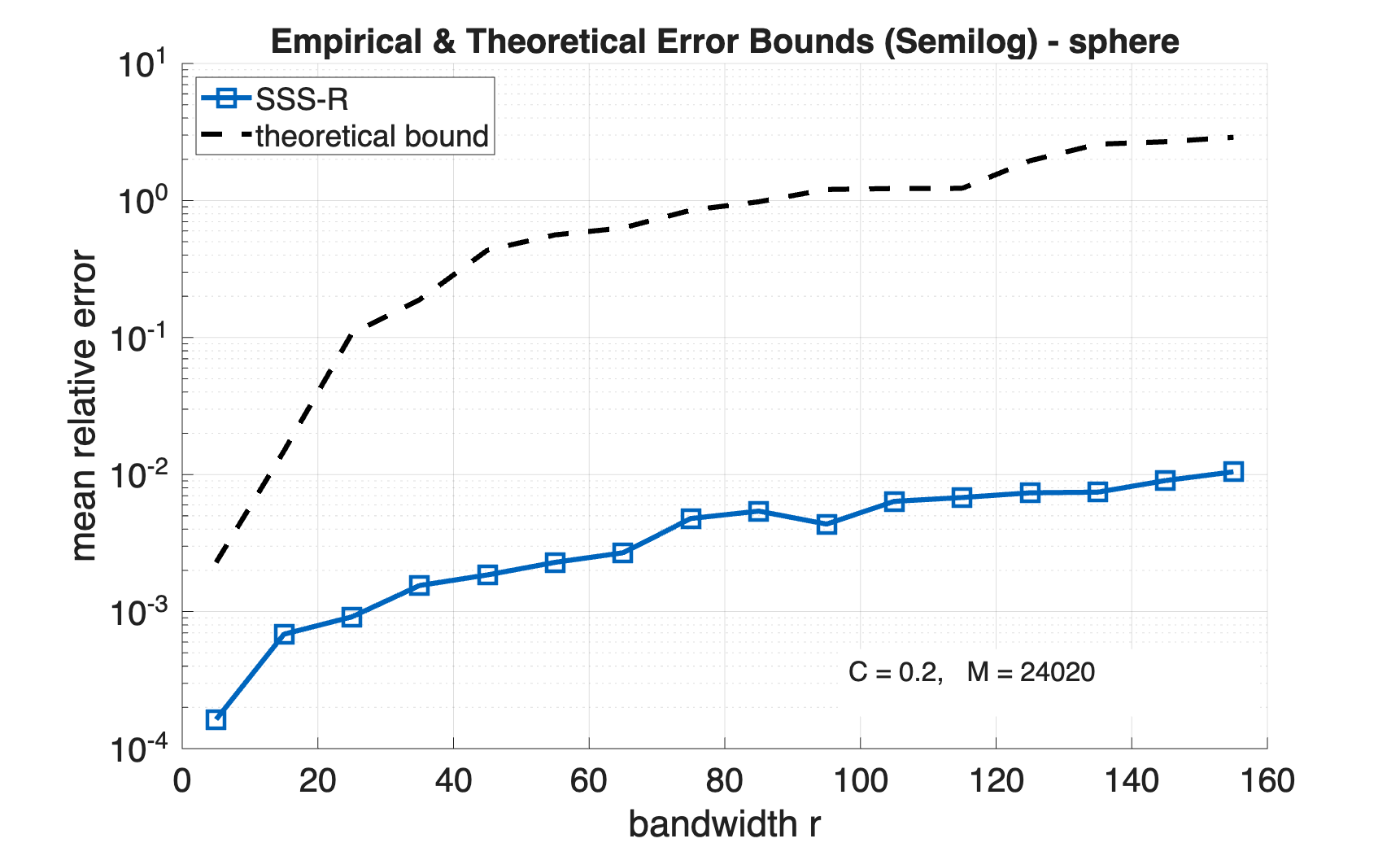}}%{Quant_error_all_1st_schemes_log_1.png}}
 \hfill
  \subcaptionbox{ Minnesota graph \label{fig4:d}}{\includegraphics[width=3.2in]{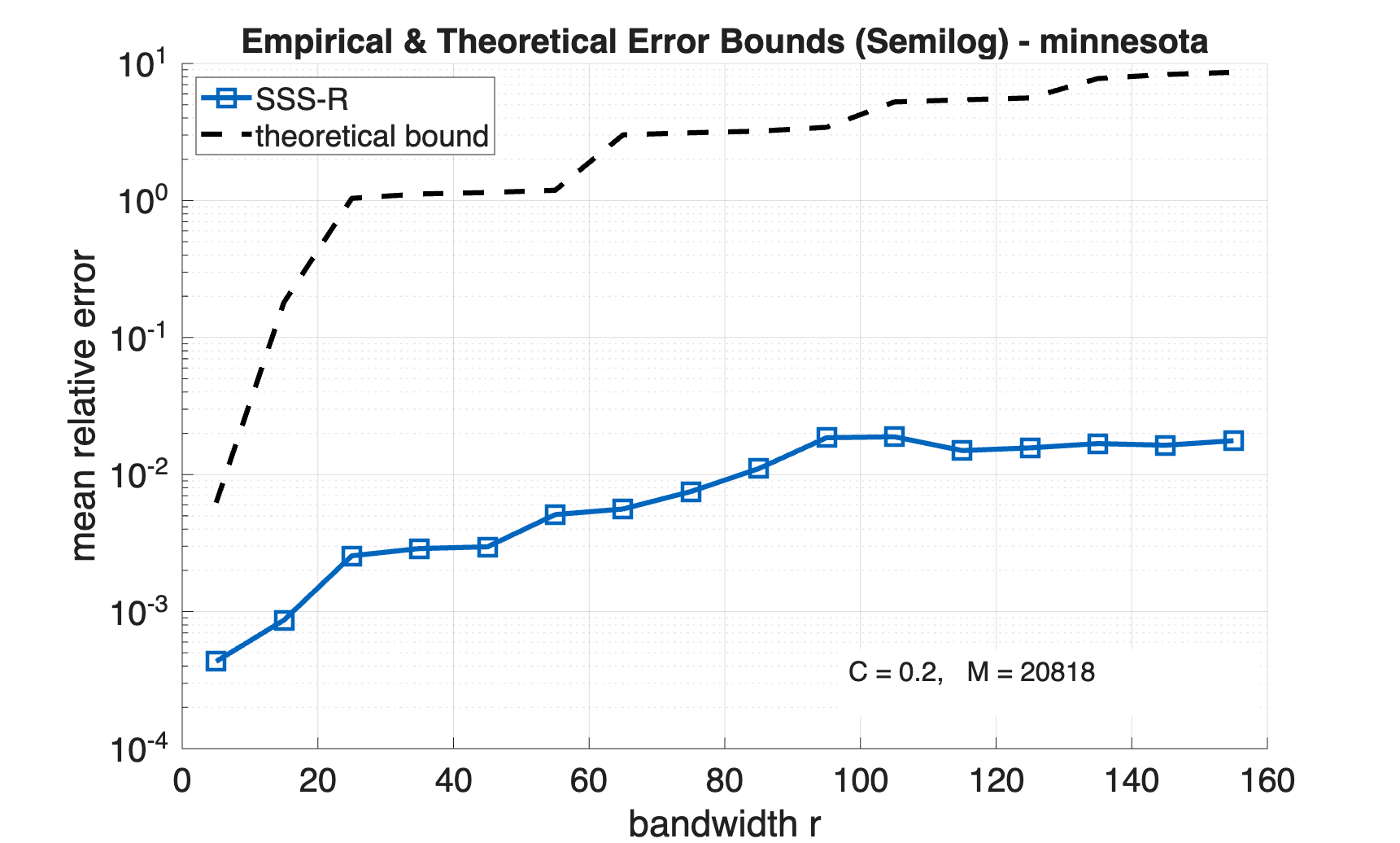}}%##

\begin{center}

   \subcaptionbox{  Swiss roll \label{fig2:a}}{\includegraphics[width=3.2in]{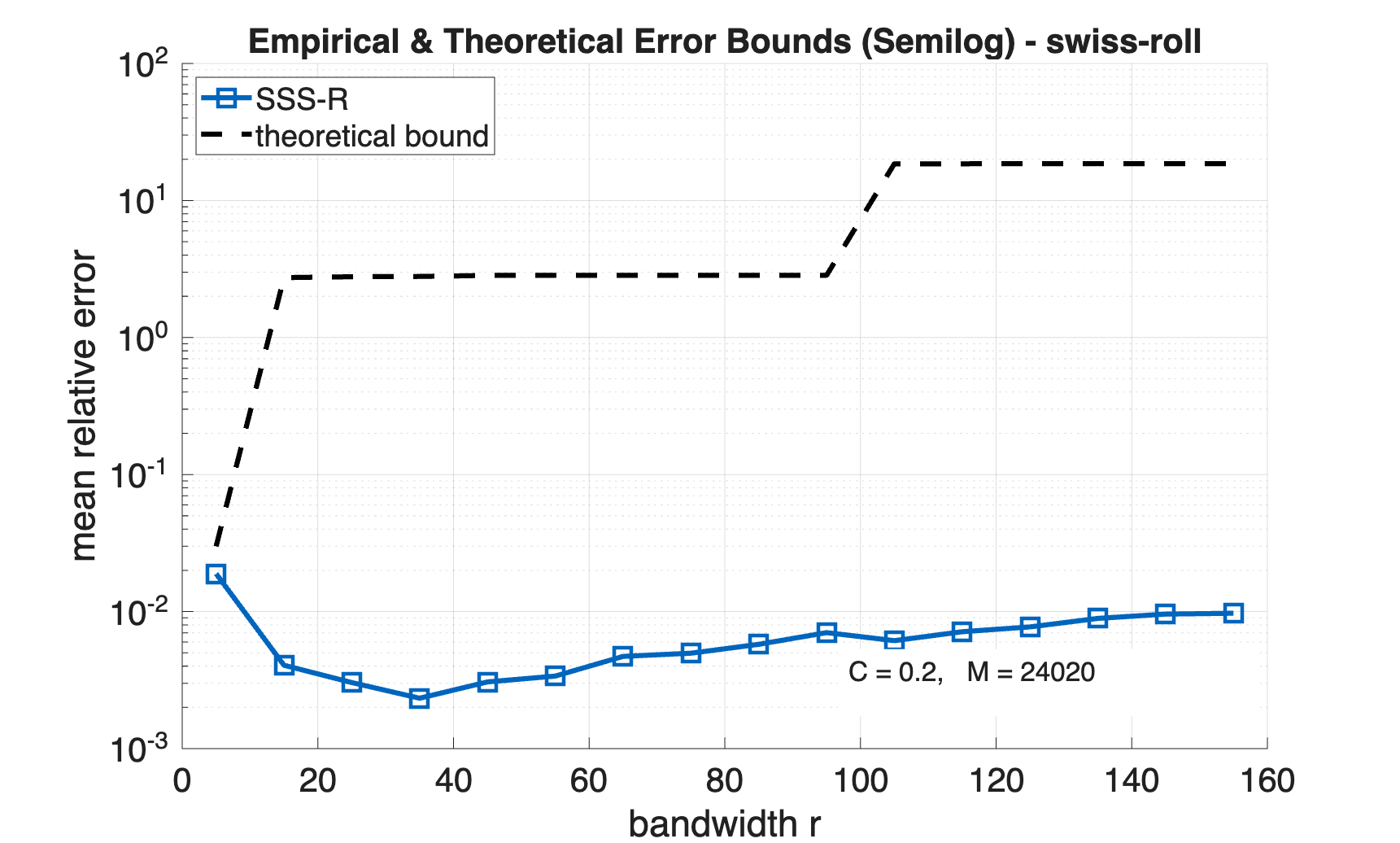}}%{Quant_error_all_1st_schemes_log_1.png}}
  \end{center}

  \caption{Illustration of theoretical and empirical error bounds of  \textit{Step-by-Step-Serving with Replacement} (SSS-R) algorithm presented in ~Algorithm~\ref{alg:NlogN-bits}. 
  }
  \label{fig:num:theory-empirics}
\end{figure}

\paragraph{Experiment 3: Validation of Theoretical Error Bounds for SSS-R.}
Figure~\ref{fig:num:theory-empirics} illustrates the empirical and theoretical mean relative error bounds of the Step-by-Step-Serving with Replacement (SSS-R) algorithm as a function of the bandwidth parameter $r$ on five different graph structures: a 2D grid graph, the Bunny graph, the Sphere graph, the Minnesota graph, and the Swiss roll dataset. The solid blue curves correspond to the empirical errors as in \eqref{eq:relat-im-error} obtained from numerical experiments, while the dashed black curves represent the theoretical error bounds derived in Algorithm~3.

Across all graph types, the empirical errors remain consistently below the corresponding theoretical bounds for the entire range of bandwidth values. This indicates that the theoretical results provide conservative worst-case guarantees, while the practical performance of SSS-R is substantially better. Nevertheless, the theoretical bounds correctly capture the qualitative dependence of the error on the bandwidth parameter.

As the bandwidth $r$ increases, the empirical error generally exhibits a monotonic increase, reflecting the increased smoothing and loss of local information associated with larger bandwidths. An exception is observed for the Swiss roll dataset, where the empirical error initially decreases for small values of $r$ before increasing, suggesting the existence of an optimal intermediate bandwidth for manifold-structured data.

Despite the gap in magnitude -- often spanning one or more orders between theory and practice -- the empirical and theoretical curves follow similar trends across all datasets. These results demonstrate that the SSS-R algorithm achieves low empirical error across a wide range of graph geometries and validates the robustness of the proposed method with respect to different underlying graph structures.

\paragraph{Experiment 4: Quantization of Approximately Bandlimited Graph Signals on Meshes.}
%\text{\color{blue} Jinna's part}
To assess the robustness of the proposed quantization methods beyond the ideal bandlimited setting that is explained by our theory, we conduct a numerical experiment that is inspired by the halftoning problem.  Using the PyVista Python package, we load the Stanford Bunny mesh and construct an undirected graph over its vertices via a k-nearest-neighbors rule, following a procedure analogous to that used in the GSPBox toolbox. The associated graph Laplacian is then computed from the resulting adjacency structure.
For that three-dimensional mesh plot, we consider an approximately bandlimited graph signal that assigns to each mesh vertex its
z-coordinate.

\begin{figure}[!htbp]
  \centering  
  % Row 1: MSQ (original + error)
  \begin{subfigure}{0.23\textwidth}
    \centering
    \includegraphics[width=1.2\linewidth]{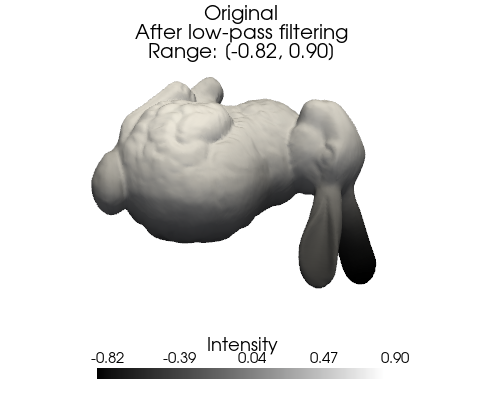}
    \caption{Original}
    \label{figmesh:a}
  \end{subfigure}
  \hfill
  \begin{subfigure}{0.23\textwidth}
    \centering
    \includegraphics[width=1.2\linewidth]{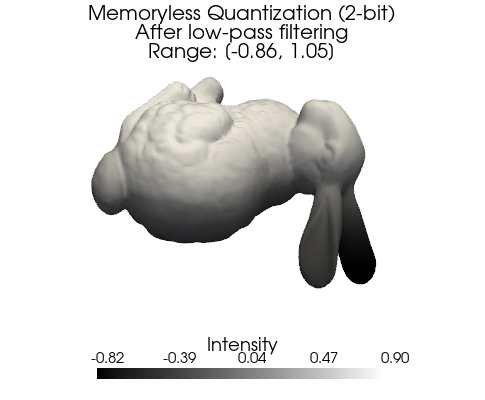}
    \caption{MSQ 2 bits }
    \label{figmesh:b}
  \end{subfigure}
  \hfill
  \begin{subfigure}{0.23\textwidth}
    \centering
    \includegraphics[width=1.2\linewidth]{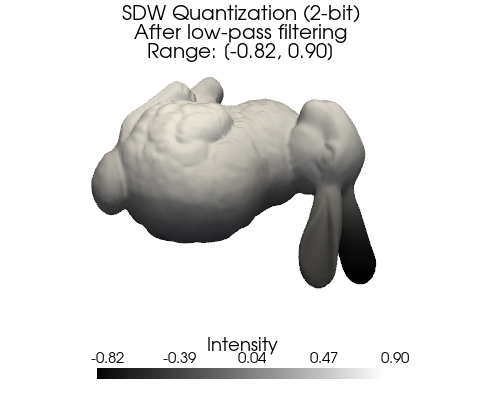}
    \caption{SDW 2 bits }
    \label{figmesh:c}
  \end{subfigure}
  \hfill
  \begin{subfigure}{0.23\textwidth}
    \centering
    \includegraphics[width=1.2\linewidth]{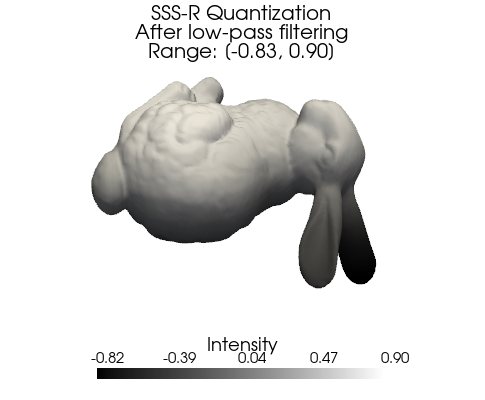}
    \caption{SSS-R 2 bits}
    \label{figmesh:d}
  \end{subfigure}
  
  \vspace{0.5cm}
  
  % Row 2: spectrum + error
  \begin{subfigure}{0.23\textwidth}
    \centering
\includegraphics[width=1.1\linewidth]{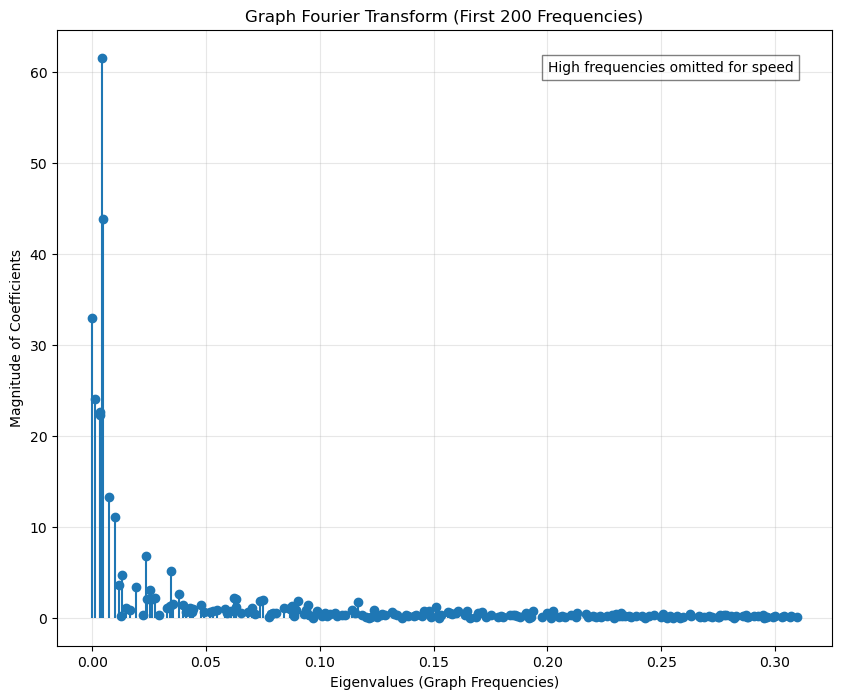}% images/TBA.jpg
    \caption{Spectrum of signal}
    \label{figmesh:e}
  \end{subfigure}
  \hfill
  \begin{subfigure}{0.23\textwidth}
    \centering
    \includegraphics[width=1.2\linewidth]{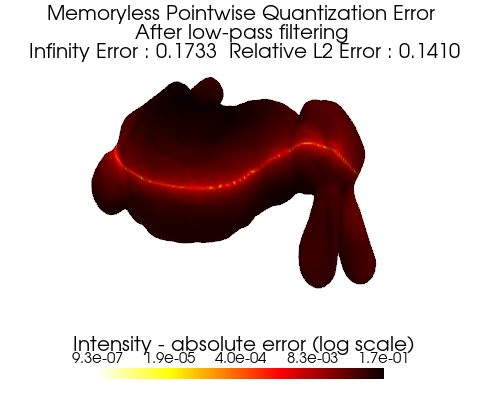}
    \caption{MSQ 2 bits error}
    \label{figmesh:f}
  \end{subfigure}
  \hfill
  \begin{subfigure}{0.23\textwidth}
    \centering
    \includegraphics[width=1.2\linewidth]{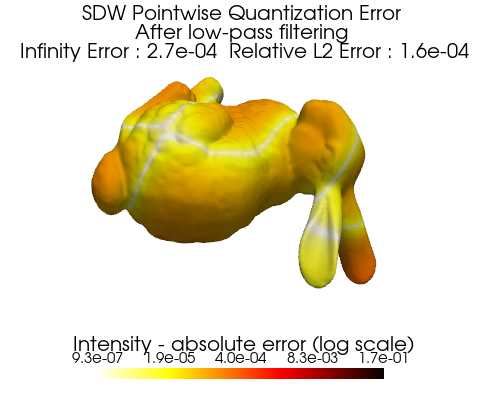}
    \caption{SDW 2 bits error}
    \label{figmesh:g}
  \end{subfigure}
  \hfill
  \begin{subfigure}{0.23\textwidth}
    \centering
    \includegraphics[width=1.2\linewidth]{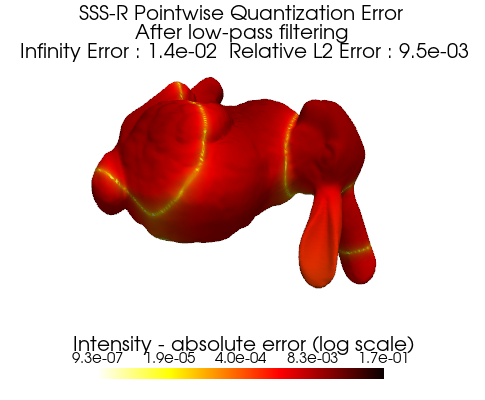}
    \caption{SSS-R 2 bits error}
    \label{figmesh:h}
  \end{subfigure}

    \vspace{0.7cm}
    
   % Row 3: without low pass filtering
  \begin{subfigure}{0.23\textwidth}
    \centering
    \includegraphics[width=1.2\linewidth]{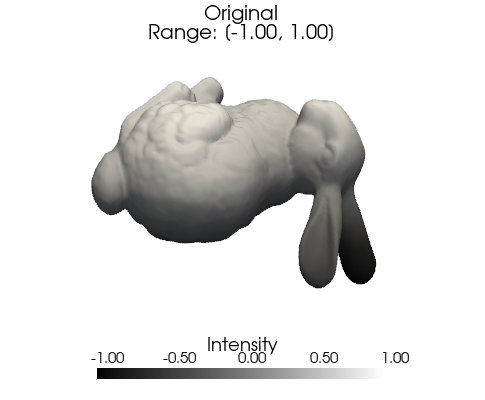}
    \caption{Original without low-pass filtering}
    \label{figmesh:i}
  \end{subfigure}
  \hfill
  \begin{subfigure}{0.23\textwidth}
    \centering
    \includegraphics[width=1.2\linewidth]{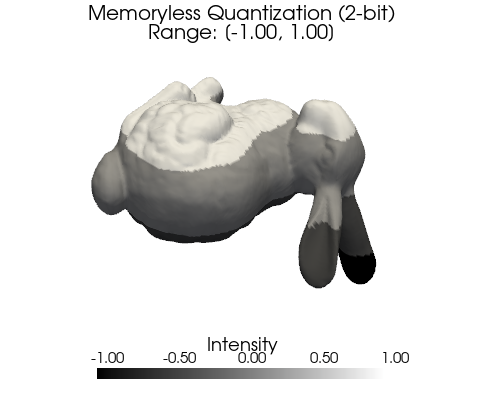}
    \caption{MSQ 2 bits without low-pass filtering}
    \label{figmesh:j}
  \end{subfigure}
  \hfill
  \begin{subfigure}{0.23\textwidth}
    \centering
    \includegraphics[width=1.2\linewidth]{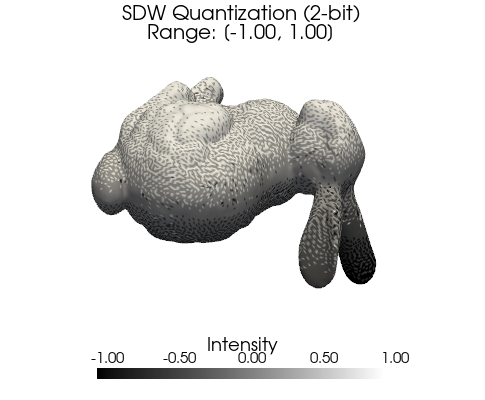}
    \caption{SDW 2 bits without low-pass filtering}
    \label{figmesh:k}
  \end{subfigure}
  \hfill
  \begin{subfigure}{0.23\textwidth}
    \centering
    \includegraphics[width=1.2\linewidth]{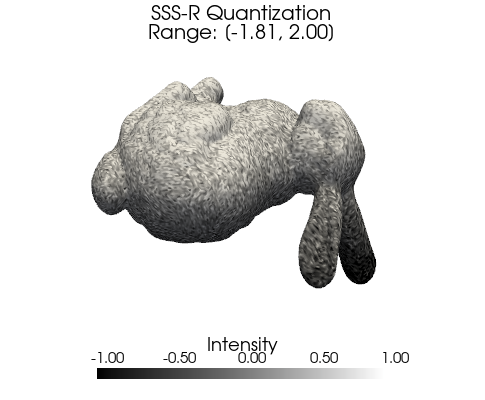}
    \caption{SSS-R 2 bits without low-pass filtering}
    \label{figmesh:l}
  \end{subfigure}
  
  \caption{Illustration of performance of the proposed quantization algorithms for a general graph data on bunny graph: Memoryless quantization, Algorithm~\ref{alg:full-with-permuations} with initialization \textit{Sigma-Delta-Weight} (SDW), and \textit{Step-by-Step-Serving with Replacement} (SSS-R) presented in Algorithm~\ref{alg:NlogN-bits}.}
  \label{fig:num:mesh}
\end{figure}

%Rather than enforcing an r-bandlimited signal, we directly use the 

%which naturally corresponds to the bunny's height and serves as
Now we quantize this non-bandlimited signal and evaluate how well the corresponding bandlimited reconstruction aligns with the low-pass part of the graph signal. This setup is motivated by key insight of digital halftoning  that human vision can be argued to correspond to a low-pass filtering operation \cite{SteinbergFloyd, lyu2023sigma, vanderhaeghe2008polyomino, krahmer2025mathematics, krahmer2022enhanced}.

%see Figure~\ref{} for a visualization using a grayscale coloring. In contrast to the theoretical setting, this signal contains energy across the entire spectrum, presenting a more challenging and realistic quantization scenario. 
We apply both memoryless quantization and the proposed graph-based quantization methods to this signal and compare their visual reconstructions as well as quantization errors on the low-pass component of the signal. 

In particular, we focus on the two best-performing methods identified in our earlier experiments on bandlimited signals: Algorithm~\ref{alg:full-with-permuations} with initialization \textit{Sigma-Delta-Weight} (SDW) and Algorithm~\ref{alg:NlogN-bits} \textit{Step-by-Step-Serving with Replacement} (SSS-R). To define the low-frequency subspace used in the noise-shaping procedures, we retain the first 20 eigenvectors of the graph Laplacian and form the corresponding low-pass projection operator. For a fair comparison, both memoryless quantization and Algorithm 1 with SDW are implemented using 2-bit (approximately $\log\log(N)$) quantization.

The resulting visual reconstructions and quantitative error metrics are shown in Fig.~\ref{figmesh:b}--\ref{figmesh:h}. 
Although the signal is not bandlimited and no theoretical error guarantees apply in this setting, the proposed algorithms consistently yield smaller quantization errors on the low-pass component -- both in the $l_\infty$ norm and in relative $l_2$ error -- than memoryless quantization. 

Moreover, the figure also illustrates the merit of the method for halftoning:  even though the original signal is not bandlimited, the noise-shaping reconstructions give rise to good visual fidelity not only to the low-pass component, but also to the original, as shown in Fig.~\ref{figmesh:i}--\ref{figmesh:l}, indicating that indeed the visual appearance is dominated by the low-pass components.

% \begin{figure}[h!]

% \begin{center}

%  %   \subcaptionbox{ MSQ 1 bit  \label{fig2:a}}{\includegraphics[width=4in]{images/bunny_memoryless_1bit.png}}%{Quant_error_all_1st_schemes_log_1.png}}
%  % \hfill
   
% \subcaptionbox{ MSQ 2bits \label{figmesh:a}}{\includegraphics[width=4in]{images/bunny_mml_knn_2bits_wlowpass.png}}%##

%  \hfill

%  \subcaptionbox{ MSQ 2bits quantization error \label{figmesh:b}}{\includegraphics[width=4in]{images/error_bunny_mml_knn_2bits_wlowpass.png}}%##

%  \hfill
 
% \subcaptionbox{  SDW 2 bits \label{figmesh:c}}{\includegraphics[width=4in]{images/bunny_sdw_knn_2bits_wlowpass.png}}%{Quant_error_all_1st_schemes_log_1.png}}

% \hfill

% \subcaptionbox{  SDW 2 bits quantization error \label{figmesh:d}}{\includegraphics[width=4in]{images/error_bunny_sdw_knn_2bits_wlowpass.png}}%

% \hfill

% \subcaptionbox{  SSS-R \label{figmesh:e}}{\includegraphics[width=4in]{images/bunny_sssr_knn_2bits_wlowpass.png}}
  
% \hfill

% \subcaptionbox{  SSS-R quantization error\label{figmesh:f}}{\includegraphics[width=4in]{images/error_bunny_sssr_knn_2bits_wlowpass.png}}
  
%   \end{center}

%   \caption{Illustration of performance of the proposed quantization algorithms for a general graph data on bunny graph: Memoryless quantiztaion, Algorithm~\ref{alg:full-with-permuations} with initialization \textit{Sigma-Delta-Weight} (SDW), and \textit{Step-by-Step-Serving with Replacement} (SSS-R) presented in Algorithm~\ref{alg:NlogN-bits}.}
%   \label{fig:num:theory-empirics}
% \end{figure}

\section{Proofs}

\subsection{Auxiliary Lemmas}
Before proving Theorem~\ref{Thm1}, we collect several auxiliary results related to properties of bandlimited graph signals and to concentration inequalities. We begin with a basic property of bandlimited graph signals.

\begin{lem}\label{lem:2norm-of-fun}
Let $\f:\V \to \R$ be an $r$-bandlimited graph signal with unit infinity norm. Then its $\ell_2$-norm is bounded from below as
\begin{equation*}
    \|\f\|_2^2 \ge \frac{N}{r\,\mu_r},
\end{equation*}
where $\mu_r$ denotes the incoherence of the subspace spanned by $\Xb_r$.
\end{lem}

\begin{proof}
Since $\f$ is $r$-bandlimited with unit infinity norm, there exists a vector $\alpha \in \R^r$ such that
$\widehat{\f}=\Xb_r \alpha$ and
\[
\f=\frac{1}{\|\widehat{\f}\|_\infty}\Xb_r \alpha .
\]
Therefore, the $\ell_2$-norm of $\f$ can be written as
\begin{equation*}
    \|\f\|_2^2=\frac{1}{\|\widehat{\f}\|_{\infty}^2}\|\widehat{\f}\|_2^2
    =\frac{\|\alpha\|_2^2}{\|\widehat{\f}\|_{\infty}^2},
\end{equation*}
where we used the orthogonality of the columns of $\Xb_r$, which implies
$\|\widehat{\f}\|_2^2=\|\alpha\|_2^2$.

Next, we bound $\|\widehat{\f}\|_\infty^2$. Let us represent $\Xb_r$ by its rows,
$\Xb_r^T=[\x^1,\ldots,\x^N]\in\R^{r\times N}$, so that
\[
\widehat{\f}_i=\lra{\x^i,\alpha}, \qquad i=1,\ldots,N.
\]
Then
\[
\|\widehat{\f}\|_\infty^2
= \max_i |\lra{\x^i,\alpha}|^2
\le \max_i \|\x^i\|_2^2 \, \|\alpha\|_2^2.
\]
By definition of incoherence,
\[
\max_i \|\x^i\|_2^2 = \frac{r}{N}\mu_r.
\]
Hence,
\[
\|\widehat{\f}\|_\infty^2 \le \mu_r \frac{r}{N}\|\alpha\|_2^2.
\]
Substituting into the expression for $\|\f\|_2^2$ yields
\[
\|\f\|_2^2
\ge \frac{\|\alpha\|_2^2}{\mu_r \frac{r}{N}\|\alpha\|_2^2}
= \frac{N}{r\mu_r},
\]
which proves the claim.
\end{proof}

\begin{lem}\label{lem:app_of_MBernstein}
Consider the extended filter matrix
$\widetilde{\Lb}=[\elb_{k_1},\elb_{k_2},\cdots,\elb_{k_M}] \in \R^{r\times M}$
and the extended signal
$\widetilde{\f}= [\f_{k_1},\f_{k_2},\cdots,\f_{k_M}]\in \R^{M}$,
where the indices $k_i$, $i=1,\ldots,M$, are selected in each iteration of Algorithm~\ref{alg:NlogN-bits}. Then, with probability at least $1-\delta$,
\begin{equation*}
     \|\widetilde \Lb  \widetilde \f - \E \widetilde \Lb \widetilde \f \|^2
     \le C\frac{M}{N}\mu r\log\!\left(\frac{r+1}{\delta}\right),
\end{equation*}
for any $0<\delta<1$ and for some absolute constant $C>0$.
\end{lem}

\begin{proof}
We apply the Matrix Bernstein inequality (see, e.g., Theorem~1.6.2 in \cite{tropp2015introduction}). For a random matrix
$\Zb\in\R^{r_1\times r_2}$, define the matrix variance statistic as
\[
v(\Zb)= \max\{\|\E \Zb \Zb^T\|,\|\E \Zb^T \Zb\|\}.
\]
Let $\Zb_1,\ldots,\Zb_M\in\R^{r_1\times r_2}$ be independent, centered random matrices such that
\[
\E \Zb_k=\bm 0, \qquad \|\Zb_k\|\le L, \quad k=1,\ldots,M.
\]
Then, for all $t\ge 0$,
\[
\Prob\!\left(\Big\|\sum_{i=1}^M\Zb_i\Big\|>t\right)
\le (r_1+r_2)\exp\!\left(\frac{-t^2/2}{v+Lt/3}\right),
\]
where $v=v(\sum_{i=1}^M\Zb_i)$.

For each $i=1,\ldots,M$, let $\Sb_i\in\R^r$ be an independent random vector uniformly distributed over the set
$\{\f_1\elb_1,\f_2\elb_2,\ldots,\f_N\elb_N\}\subset\R^r$.
Define
\[
\Zb_i=\Sb_i-\E \Sb_i
= \Sb_i -\frac{1}{N}\sum_{k=1}^N \f_k \elb_k .
\]
Using the incoherence bound, we obtain almost surely
\[
\|\Zb_i\|
\le \frac{1}{N}\sum_{k\ne k_i}\|\f_k \elb_k\|
\le \frac{N-1}{N}\sqrt{\frac{\mu r}{N}}
\le \sqrt{\frac{\mu r}{N}}.
\]

Next, we compute the matrix variance statistic. For each $i$,
\begin{align*}
\E(\Zb_i\Zb_i^T)
&= \frac{1}{N}\sum_{k=1}^N \f_k^2 \elb_k \elb_k^T
   - \frac{1}{N^2}\sum_{k,m=1}^N \f_k\f_m\elb_k\elb_m^T.
\end{align*}
Using incoherence again, we obtain
\[
\|\E(\Zb_i\Zb_i^T)\| \le \mu \frac{r}{N},
\qquad
\|\E(\Zb_i^T\Zb_i)\| \le \mu \frac{r}{N}.
\]
Therefore,
\begin{align*}
v\!\left(\sum_{i=1}^M\Zb_i\right)
&= \max\!\left\{
\Big\|\sum_{i=1}^M \E(\Zb_i\Zb_i^T)\Big\|,
\Big\|\sum_{i=1}^M \E(\Zb_i^T\Zb_i)\Big\|
\right\} \\
&\le \frac{M}{N}\mu r .
\end{align*}

Applying Matrix Bernstein yields
\[
\Prob \big(\|\widetilde \Lb  \widetilde \f - \E \widetilde \Lb \widetilde \f\|>t\big)
\le (r+1) \exp\!\left(-c \min \Big\{ t^2 \tfrac{N}{M\mu r},\;
t \sqrt{\tfrac{N}{\mu r}} \Big\}\right).
\]
Setting
\[
t=\sqrt{C\frac{M}{N}\mu r\log\!\left(\frac{r+1}{\delta}\right)}
\]
gives, with probability at least $1-\delta$,
\[
\|\widetilde \Lb  \widetilde \f - \E \widetilde \Lb \widetilde \f\|^2
\le C\frac{M}{N}\mu r\log\!\left(\frac{r+1}{\delta}\right),
\]
which completes the proof.
\end{proof}

\subsection{Proof of Theorem~\ref{Thm1}}
Let
$\widetilde{\Lb}=[\elb_{k_1},\elb_{k_2},\cdots,\elb_{k_M}] \in \R^{r\times M}$
be the extended filter matrix formed from Algorithm~\ref{alg:NlogN-bits}, and let
$\widetilde{\f}= [\f_{k_1},\f_{k_2},\cdots,\f_{k_M}]\in \R^{M}$ be the corresponding extended signal.

The proof consists of two parts. First, we show that $\widetilde{\f}$ and its quantized version $\widetilde{\q}$ are close under the action of $\widetilde{\Lb}$. Second, we relate $\widetilde{\Lb}\widetilde{\f}$ to the true signal $\f$ and derive the final quantization error bound.

At iteration $i$ of Algorithm~\ref{alg:NlogN-bits}, let $\elb_{k_i}$ denote the selected column of $\Xb_r^T$, and let $\uu_i$ be the corresponding state variable. By incoherence,
$\|\elb_{k_i}\|^2 \le \mu \frac{r}{N}$.

We distinguish two cases depending on whether
\[
\f_{k_i}+ \frac{\lra{\elb_{k_i},\uu_{i-1}}}{\|\elb_{k_i}\|^2}
\]
lies within the alphabet range.

\textit{Case 1: within the alphabet.}
If
\[
\Big|\f_{k_i}+ \frac{\lra{\elb_{k_i},\uu_{i-1}}}{\|\elb_{k_i}\|^2}\Big|\le q_{\max},
\]
then for
\[
\widetilde \q_i=Q_{\delta}\!\left( \f_{k_i}+ \frac{\lra{\elb_{k_i},\uu_{i-1}}}{\|\elb_{k_i}\|^2} \right),
\]
we have
\[
\Big|\widetilde \q_i-\Big(\f_{k_i}+ \frac{\lra{\elb_{k_i},\uu_{i-1}}}{\|\elb_{k_i}\|^2}\Big)\Big|
\le \frac{\delta}{2}.
\]
Let $P_{\elb_{k_i}}$ denote the projection onto $\mathrm{span}\{\elb_{k_i}\}$. Then
\[
P_{\elb_{k_i}}\uu_i
=\elb_{k_i}\Big( \frac{\elb_{k_i}^T \uu_{i-1}}{\|\elb_{k_i}\|^2}
+\f_{k_i}-\widetilde \q_i\Big),
\]
and hence
\begin{equation}\label{eq:bound-for-projection}
\| P_{\elb_{k_i}}\uu_i\|^2
\le \|\elb_{k_i}\|^2 \frac{\delta^2}{4}
\le \frac{\mu\delta^2 r}{4N}.
\end{equation}
Therefore,
\[
\|\uu_i\|^2
\le \frac{\mu\delta^2 r}{4N} + \|\uu_{i-1}\|^2 .
\]

\textit{Case 2: outside the alphabet.}
If
\[
\Big|\f_{k_i}+ \frac{\lra{\elb_{k_i},\uu_{i-1}}}{\|\elb_{k_i}\|^2}\Big|> q_{\max},
\]
assume without loss of generality that the expression exceeds $q_{\max}$. Then $\widetilde \q_i=q_{\max}$ and
\[
\Big|\f_{k_i}+ \frac{\lra{\elb_{k_i},\uu_{i-1}}}{\|\elb_{k_i}\|^2}-\widetilde \q_i\Big|
\le \Big|\frac{\lra{\elb_{k_i},\uu_{i-1}}}{\|\elb_{k_i}\|^2}\Big|.
\]
Consequently,
\[
\|P_{\elb_{k_i}}\uu_i\|
\le \|P_{\elb_{k_i}}\uu_{i-1}\|,
\]
and therefore $\|\uu_i\|^2 \le \|\uu_{i-1}\|^2$.

At initialization,
\[
\|\uu_1\|^2=\|\elb_{k_1}\|^2|\f_{k_1}-\widetilde \q_1|^2
\le \mu \frac{\delta^2 r}{4N}.
\]
Combining both cases yields
\[
\|\uu_M\|^2 \le \|\uu_1\|^2 + M\frac{\mu\delta^2 r}{4N}
\le \mu\delta^2 r\frac{1+M}{N}.
\]
Thus,
\begin{equation}\label{eq:1st-bound}
\|\widetilde{\Lb}(\widetilde{\f}-\widetilde \q)\|_2^2
=\|\uu_M\|^2
\le C_1 \mu\delta^2 \frac{rM}{N}
\le C_2 \mu r \frac{M}{N}.
\end{equation}

Next, observe that
\[
\E\widetilde{\Lb}\widetilde{\f}
= \E \sum_{i=1}^M \elb_{k_i}\f_{k_i}
= \frac{M}{N}\Xb_r^T\f .
\]
Applying Lemma~\ref{lem:app_of_MBernstein}, with probability at least $1-\delta$,
\begin{equation}\label{eq:2nd-bound}
\|\widetilde \Lb  \widetilde \f - \tfrac{M}{N}\Xb_r^T\f\|^2
\le C\frac{M}{N}\mu r\log\!\left(\frac{r+1}{\delta}\right).
\end{equation}

Combining \eqref{eq:1st-bound} and \eqref{eq:2nd-bound}, we obtain
\[
\|\widetilde{\Lb} \widetilde \q-\tfrac{M}{N} \Xb_r^T\f\|
\le \sqrt{C_2\mu\frac{rM}{N}\log\!\left(\frac{r+1}{\delta}\right)}.
\]
Rearranging terms gives the final quantization error bound
\[
\|\f-\tfrac{N}{M}\Xb_r\widetilde{\Lb}\q\|^2
\le C_2\frac{N}{M}\mu r\log\!\left(\frac{r+1}{\delta}\right),
\]
which holds with probability at least $1-\delta$.

\section{Conclusion}
We presented iterative noise-shaping methods for low-bit quantization of bandlimited graph signals. 
Permutation-based methods with structured initialization and random sampling methods with theoretical guarantees both achieve low reconstruction error. 
Future work includes extending to time-varying graph signals and adaptive quantization schemes.

\printbibliography

@inproceedings{hassibi2002expected,
  title={On the expected complexity of integer least-squares problems},
  author={Hassibi, Babak and Vikalo, Haris},
  booktitle={2002 IEEE International Conference on Acoustics, Speech, and Signal Processing},
  volume={2},
  pages={II--1497},
  year={2002},
  organization={IEEE}
}

@article{chou2015noise,
  title={Noise-shaping quantization methods for frame-based and compressive sampling systems},
  author={Chou, Evan and G{\"u}nt{\"u}rk, C Sinan and Krahmer, Felix and Saab, Rayan and Y{\i}lmaz, {\"O}zg{\"u}r},
  journal={Sampling Theory, a Renaissance: Compressive Sensing and Other Developments},
  pages={157--184},
  year={2015},
  publisher={Springer}
}

@article{saab2018quantization,
  title={Quantization of compressive samples with stable and robust recovery},
  author={Saab, Rayan and Wang, Rongrong and Y{\i}lmaz, {\"O}zg{\"u}r},
  journal={Applied and Computational Harmonic Analysis},
  volume={44},
  number={1},
  pages={123--143},
  year={2018},
  publisher={Elsevier}
}

@article{maly2022simple,
  title={A simple approach for quantizing neural networks},
  author={Maly, Johannes and Saab, Rayan},
  journal={arXiv preprint arXiv:2209.03487},
  year={2022}
}

@inproceedings{cui2020adaptive,
  title={Adaptive graph encoder for attributed graph embedding},
  author={Cui, Ganqu and Zhou, Jie and Yang, Cheng and Liu, Zhiyuan},
  booktitle={Proceedings of the 26th ACM SIGKDD international conference on knowledge discovery \& data mining},
  pages={976--985},
  year={2020}
}

@inproceedings{rehman2010flicker,
  title={Flicker assessment of low-to-medium frame-rate binary video halftones},
  author={Rehman, Hamood-Ur and Evans, Brian L},
  booktitle={2010 IEEE Southwest Symposium on Image Analysis \& Interpretation (SSIAI)},
  pages={185--188},
  year={2010},
  organization={IEEE}
}

@article{gunturk2021approximation,
  title={Approximation of functions with one-bit neural networks},
  author={G{\"u}nt{\"u}rk, C Sinan and Li, Weilin},
  journal={arXiv preprint arXiv:2112.09181},
  year={2021}
}

@article{krahmer2012root,
  title={Root-exponential accuracy for coarse quantization of finite frame expansions},
  author={Krahmer, Felix and Saab, Rayan and Ward, Rachel},
  journal={IEEE transactions on information theory},
  volume={58},
  number={2},
  pages={1069--1079},
  year={2012},
  publisher={IEEE}
}

@article{gunturk2013sobolev,
  title={Sobolev duals for random frames and Sigma--Delta quantization of compressed sensing measurements},
  author={G{\"u}nt{\"u}rk, C Sinan and Lammers, Mark and Powell, Alexander M and Saab, Rayan and Y{\i}lmaz, {\"O}},
  journal={Foundations of Computational mathematics},
  volume={13},
  pages={1--36},
  year={2013},
  publisher={Springer}
}

@article{daubechies2003approximating,
  title={Approximating a bandlimited function using very coarsely quantized data: A family of stable sigma-delta modulators of arbitrary order},
  author={Daubechies, Ingrid and DeVore, Ron},
  journal={Annals of mathematics},
  volume={158},
  number={2},
  pages={679--710},
  year={2003},
  publisher={JSTOR}
}

@article{gunturk2003one,
  title={One-bit sigma-delta quantization with exponential accuracy},
  author={G{\"u}nt{\"u}rk, C Sinan},
  journal={Communications on Pure and Applied Mathematics: A Journal Issued by the Courant Institute of Mathematical Sciences},
  volume={56},
  number={11},
  pages={1608--1630},
  year={2003},
  publisher={Wiley Online Library}
}

@article{perraudin2014gspbox,
  title={GSPBOX: A toolbox for signal processing on graphs},
  author={Perraudin, Nathana{\"e}l and Paratte, Johan and Shuman, David and Martin, Lionel and Kalofolias, Vassilis and Vandergheynst, Pierre and Hammond, David K},
  journal={arXiv preprint arXiv:1408.5781},
  year={2014}
}

@article{lybrand2021greedy,
  title={A greedy algorithm for quantizing neural networks},
  author={Lybrand, Eric and Saab, Rayan},
  journal={The Journal of Machine Learning Research},
  volume={22},
  number={1},
  pages={7007--7044},
  year={2021},
  publisher={JMLRORG}
}

@article{brooks2013non,
  title={Non-localization of eigenfunctions on large regular graphs},
  author={Brooks, Shimon and Lindenstrauss, Elon},
  journal={Israel Journal of Mathematics},
  volume={193},
  number={1},
  pages={1--14},
  year={2013},
  publisher={Springer}
}

@article{dekel2011eigenvectors,
  title={Eigenvectors of random graphs: Nodal domains},
  author={Dekel, Yael and Lee, James R and Linial, Nathan},
  journal={Random Structures \& Algorithms},
  volume={39},
  number={1},
  pages={39--58},
  year={2011},
  publisher={Wiley Online Library}
}

@article{lyu2020sigma,
  title={Sigma Delta quantization for images},
  author={Lyu, He and Wang, Rongrong},
  journal={arXiv preprint arXiv:2005.08487},
  year={2020}
}

@article{krahmer2022enhanced,
  title={Enhanced Digital Halftoning via Weighted Sigma-Delta Modulation},
  author={Krahmer, Felix and Veselovska, Anna},
  journal={arXiv preprint arXiv:2202.04986},
  year={2022}
}

@article{tropp2015introduction,
  title={An introduction to matrix concentration inequalities},
  author={Tropp, Joel A and others},
  journal={Foundations and Trends{\textregistered} in Machine Learning},
  volume={8},
  number={1-2},
  pages={1--230},
  year={2015},
  publisher={Now Publishers, Inc.}
}

@inproceedings{krahmer2023quantization, title={Quantization of bandlimited graph signals}, author={Krahmer, Felix and Lyu, He and Saab, Rayan and Veselovska, Anna and Wang, Rongrong}, booktitle={2023 International Conference on Sampling Theory and Applications (SampTA)}, pages={1--5}, year={2023}, organization={IEEE} }

@article{lyu2023sigma,
  title={Sigma Delta quantization for images},
  author={Lyu, He and Wang, Rongrong},
  journal={Communications on Pure and Applied Mathematics},
  volume={76},
  number={5},
  pages={901--945},
  year={2023},
  publisher={Wiley Online Library}
}

@article{vanderhaeghe2008polyomino,
  title={Polyomino-based digital halftoning},
  author={Vanderhaeghe, David and Ostromoukhov, Victor},
  journal={arXiv preprint arXiv:0812.1647},
  year={2008}
}

@article{shuman2016vertex,
  title={Vertex-frequency analysis on graphs},
  author={Shuman, David I and Ricaud, Benjamin and Vandergheynst, Pierre},
  journal={Applied and Computational Harmonic Analysis},
  volume={40},
  number={2},
  pages={260--291},
  year={2016},
  publisher={Elsevier}
}

@article{krahmer2025mathematics,
  title={The mathematics of dots and pixels: On the theoretical foundations of image halftoning},
  author={Krahmer, Felix and Veselovska, Anna},
  journal={GAMM-Mitteilungen},
  volume={48},
  number={1},
  pages={e70000},
  year={2025},
  publisher={Wiley Online Library}
}

@article{SteinbergFloyd,
  author    = {Floyd, R. W. and Steinberg, L.},
  title     = {An Adaptive Algorithm for Spatial Grey Scale},
  journal   = {Proceedings of the Society of Information Display},
  volume    = {17},
  year      = {1976},
  pages     = {75-77}
}

@article{pesenson2009variational,
  title={Variational splines and Paley–Wiener spaces on combinatorial graphs},
  author={Pesenson, Isaac},
  journal={Constructive Approximation},
  year={2009}
}

@article{zhang2022post,
  title={Post‑training quantization for deep learning},
  author={Zhang, Xiang and et al.},
  year={2022}
}

@book{gray1998quantization,
  title={Quantization},
  author={Gray, Robert M. and Neuhoff, David L.},
  year={1998}
}

\end{document}